\documentclass[11pt]{article}

\usepackage{amsmath,amsfonts,axodraw}

\def\bea{\begin{eqnarray}}
\def\eea{\end{eqnarray}}
\def\be{\begin{equation}}

\def\Z{{\bf Z}}
\newcommand{\Tr}{\text{Tr}}

\begin{document}

%%%%%%%%%%
%%%%%%%%%%      title page
%%%%%%%%%%

\begin{flushright} CERN--TH/2002--046 \end{flushright}
\vspace{1.5mm}
\vspace{1.5cm}
\begin{center}

\def\thefootnote{\fnsymbol{footnote}}

{\Large \bf Anomalies, Fayet--Iliopoulos terms and the \\[0.2cm] 
 consistency of orbifold field theories} \\[1cm]
{\large R. Barbieri$^1$, R. Contino$^{1,2}$, P. Creminelli$^{1,2}$, \\[0.5cm]
 R. Rattazzi$^{2\,}$\footnote{On leave from INFN, Pisa, Italy.}, C.A. Scrucca$^2$}
\\[1.5cm]

{\small 
$^1$\textit{Scuola Normale Superiore, Piazza dei Cavalieri 7, I-56126 Pisa, $\&$ INFN,
Italy} 
\\[0.3cm] $^2$\textit{Theory Division, CERN, CH-1211 Geneva 23, Switzerland} }

\end{center}

\vspace{2cm}

\hrule 
\vspace{0.5cm} 
{\small  \noindent \textbf{Abstract} \\[0.3cm]
\noindent
We study the consistency of orbifold field theories and clarify to what extent
the condition of having an anomaly-free  spectrum of zero-modes is sufficient
to guarantee it.  Preservation of gauge invariance at the quantum
level is possible, although at the price, in general, of introducing
operators that break the 5d local parity. These operators are,
however, perfectly consistent  with the orbifold
projection. We also clarify the relation between localized
Fayet--Iliopoulos (FI) terms and anomalies. These terms can be consistently added,
breaking neither local supersymmetry nor gauge symmetry. 
In the framework of supergravity the localized FI term arises as the boundary completion of a
bulk interaction term: given the bulk Lagrangian the FI is fixed by gauge invariance.
}
\vspace{0.5cm}  
\hrule

\def\thefootnote{\arabic{footnote}}
\setcounter{footnote}{0}

\newpage

%%%%%%%%%%
%%%%%%%%%%      main part 
%%%%%%%%%%

\section{Introduction}

The existence of extra compact dimensions has been recognized 
since long ago to be quite a natural and appealing possibility,
subject to rather mild constraints from present-day experiments.
String theory, which is the only consistent candidate available 
at present for a microscopic and unified description of fundamental 
interactions, requires in fact at least six additional space dimensions; 
a fact that suggests that these might play some important role also from an 
effective field theory point of view~\cite{Antoniadis:1990ew}.

Recently, renewed attention has been devoted to orbifold field 
theories with Scherk--Schwarz (SS) symmetry breaking \cite{Scherk:1978tazr}. 
Starting from five-dimensional (5d) supersymmetric theories, for instance, 
it has been possible to construct very simple and interesting extensions 
of the standard model with SS supersymmetry breaking 
\cite{Pomarol,Barbieri:2000vh}, as well as grand-unified 
theories with SS gauge symmetry breaking 
\cite{ssgauge}.
Being non-renormalizable, these models must be thought of as 
effective field theories valid up to some physical cut-off scale $\Lambda$.
Incalculable UV effects can then be parametrized by writing the most general Lagrangian
containing operators of arbitrarily high dimensions and with coefficients scaling
with the suitable inverse powers of $\Lambda$. Working at energies $E\ll \Lambda$,
and at finite order in an expansion in $E/\Lambda$ there are only a finite number of relevant
operators. It is in this weaker sense that these theories are predictive.
This is in complete analogy with the well-known case of pion interactions below the QCD
scale. One relevant energy scale of these models is given by the compactification
radius $E=1/R$. Several interesting quantities of the low energy 4d theory can
be written as a power series in $1/R\Lambda$ and ``weak'' predictivity
is satisfied only for $R\gg 1/\Lambda$. Provided this condition holds, there are
indeed some quantities which are nicely predicted in these models. These are the quantities for
which there exists some 5d symmetry that forbids direct contributions by local
5d operators but is broken by the compactification. This
ensures that the value of these quantities is controlled by calculable IR effects,
and can thus be expressed in terms of other observables
(see \cite{Antoniadis:1998sd,Barbieri:2000vh,Arkani-Hamed:2001mi,finiteness}).
In particular the Higgs potential is protected by the original 
$N=2$ supersymmetry and falls in this category. Then electroweak 
symmetry breaking (ESB) occurs  radiatively and is triggered by 
(global) supersymmetry breaking. Even more interestingly,  very 
constrained models can be constructed with a single light Higgs scalar before ESB, such as 
the one proposed in \cite{Barbieri:2000vh}, evading the need for at least two Higgs 
doublets in 4d supersymmetric extensions of the standard model. 
However, doubts were raised on the predictive power \cite{Ghilencea:2001bw} and even
on the consistency \cite{Scrucca:2001eb} of these models. The aim of this paper 
is to readdress these issues and discuss the conditions under which these
models can be consistently defined.

In general, orbifold compactifications do not preserve the consistency of the
higher-dimensional theory, because they correspond to a singular geometry.
The consistency of orbifold models can however be studied from an effective field
theory point of view, without invoking a more fundamental theory. 
In string orbifold models \cite{Dixon:jwjc} (see \cite{ssstringa,ssfreely} 
for specific examples with SS symmetry breaking), consistency is instead guaranteed
as a consequence of more general principles, such as modular invariance. The twisted 
sectors arising at the fixed points provide blowing-up modes which allow to resolve 
the singularities through their vacuum expectation value (VEV); orbifold compactifications 
can then be interpreted as singular limits of smooth manifold compactifications and 
this is sufficient to exclude any source of inconsistency. 
From a low energy supergravity point of view, anomalies cancel thanks to the presence
of the twisted states and a generalization \cite{Dine:xk} of the 
Green--Schwarz (GS) mechanism \cite{Green:sg} involving axions.

From an effective field theory view-point, one might think that a simple sufficient 
criterion to get a consistent orbifold theory could be to require an anomaly-free 
spectrum of chiral zero modes. 
This has been shown to hold true in the simplest case of $S^1/\Z_2$ orbifolds
\cite{Arkani-Hamed:2001is}, but needs a qualification for
$S^1/(\Z_2\times \Z_2^\prime)$
orbifolds, where anomalies may arise at fixed points and compensate
each other only globally \cite{Scrucca:2001eb}. However, as already 
argued in \cite{Scrucca:2001eb}, it is important to define the fate of the other
symmetries in the theory in order to reach any sensible conclusion.
The correct general statement turns out to be the following: under the assumption 
that the spectrum of massless modes is not anomalous, the theory can be 
made consistent and free of any localized anomaly, but in general at the price 
of introducing operators that are odd under a local 5d parity.
The reason is that there is no regularization preserving simultaneously gauge invariance 
and the 5d local parity. Notice that
these operators do not break the orbifold symmetry. As usual, the choices of preserving 
one or the other of the symmetries differ by the addition of a local counterterm that violates both 
symmetries; the localized anomalies arising in a regularization that forbids
odd operators will be canceled thanks to an induced odd Chern--Simons term in a 
gauge-invariant regularization. The connection between gauge anomalies and parity breaking by
a Chern--Simons term is indeed a well-known property of field theories in odd dimensions. 

In supersymmetric theories, another important issue concerns the generation of 
Fayet--Iliopoulos (FI) terms for the $U(1)$ gauge factors \cite{fay}. In 4d theories, this kind of 
term is not compatible with a supersymmetric and gauge-invariant vacuum state; it is 
radiatively generated only in the presence of $U(1)$-gravitational anomalies. 
Indeed, in supergravity a FI term is not gauge-invariant unless it is associated to an 
$R$-symmetry so that the gravitino has a charge \cite{Barbieri:1982ac,Ferrara:1983dh}. 
In 5d orbifold models, we find that the situation is substantially different and 
two different kinds of FI terms can arise. The first type is truly associated to anomalies
and cannot appear if the theory is consistently defined through a gauge-invariant 
regularization. The second type corresponds to the boundary completion of a bulk term
present in 5d supergravity. This bulk term is associated to, among other things, a kinetic
mixing between the $U(1)$ photon and the graviphoton.
The localized FI term is necessary to ensure full (supersymmetric) gauge invariance
of the bulk term, and as such its value at the various boundaries is fixed. 

The implication of our results for the single Higgs model of \cite{Barbieri:2000vh}
is that the theory can be made fully consistent, but at the price of introducing odd operators. 
Divergent FI terms still arise at the fixed points, unless a specific regularization is assumed.
For two Higgs models, the situation is different and one can consistently define these 
theories without the need to include odd operators.

The structure of the paper is the following. In section \ref{odd} we discuss the consistency of
odd operators in orbifold models and describe the local 5d parity that they violate. Our main discussion 
about anomalies is contained in section \ref{pv}, while we leave for the appendices the explicit description
of the regularizing procedures and an alternative approach based on Fujikawa's method. Section \ref{FI} gives
a detailed discussion about the consistency of FI terms at the orbifold fixed points and their
relation with anomalies. Finally, in section \ref{SUGRA} we discuss the compatibility of our
results with supergravity. Conclusions are drawn in section \ref{conclusions}. 

\section{\label{odd} Odd operators on orbifolds}

Consider a 5d model in which one of the spatial dimensions is compactified on a circle $S^1$, 
with coordinate $y \in [0,2\pi R]$. Consider also 5d parity $P_5$, corresponding to
a reflection of the fifth coordinate around a fixed point, say $y=0$, $y \rightarrow -y$,
leaving the other coordinates unaffected. We can choose a basis where each field $\phi$
is either even or odd: $P_5[\phi(y)]=\eta_\phi\phi(-y)$, with $\eta_\phi=\pm 1$.
One possible use of $P_5$ is to make it a (global) symmetry of the model by
forbidding $P_5$-odd operators in the Lagrangian.
Another, distinct possibility is to make it a gauge symmetry
by projecting out of our original model all
field {\it configurations} that do not satisfy the condition 
\begin{equation}
\label{eq:gauging}
\phi(y)=P_5[\phi(y)]= \eta_\phi \,\phi(-y) \;.
\end{equation}
This projection corresponds to the construction of an $S^1/\Z_2$ orbifold theory.
Notice that orbifolding corresponds to making the fields configuration $P_5$ invariant, without any reference
to the Lagrangian.
Indeed, after the projection, the addition of a parity-odd operator $O^-$ to the Lagrangian 
has no consequence because its integral over the fifth dimension trivially vanishes. 
Nevertheless, we can now add an odd operator, but with a coefficient 
that changes sign at the orbifold fixed points and is proportional to: 
\begin{equation}
c(y) = \begin{cases}
+1 & y \in (0,\pi R) \\
-1 & y \in (\pi R, 2\pi R)
\end{cases} \;.
\end{equation}
It is  obvious that the introduction of
parity-odd operators does not break the orbifold symmetry. Being a gauge symmetry the orbifold
symmetry cannot be broken: non-invariant states are simply out of the theory. Indeed if we
directly define the theory on the segment, with proper boundary conditions, we see that the idea of breaking
the orbifold symmetry is a nonsense. In the same way a non-zero profile for an odd field does not
spontaneously break $P_5$, because of eq.~(\ref{eq:gauging}).

Having realized that parity-odd operators are allowed by the orbifolding, one may ask whether it is consistent to leave 
them out, or, in other words, if they can be forbidden by another symmetry. Always remaining in the simplest
example of $S^1/\Z_2$, it is easy to see that the symmetry of reflection around the midpoint of the 
fundamental domain,
\begin{equation}
\label{eq:newparity}
\phi(y) \rightarrow  \eta \,\phi(\pi R - y) \;,
\end{equation}
forbids all parity-odd operators: with respect to (\ref{eq:gauging}), we have just changed the 
reflection point. Note that this symmetry exchanges the two boundaries: this gives no problem 
as all the fields have the same boundary conditions at $0$ and $\pi R$. This discrete global 
symmetry is precisely the $\Z^\prime_2$ parity that will be gauged in $S^1/(\Z_2 \times \Z^\prime_2)$ models. 
This symmetry is obviously explicitly broken by the introduction of odd terms and it is 
spontaneously broken by an odd profile $\langle\phi^-(y)\rangle$.  
 
In the case of an $S^1/(\Z_2 \times \Z^\prime_2)$ theory, the situation is slightly more involved. 
One may naively imagine to trivially extend the symmetry defined on $S^1/\Z_2$: reflection around 
the midpoint $\pi R/4$ of the fundamental domain. In this case we can choose one of the two possible 
parity assignments $\eta$ and $\eta'$ given by the two orbifold actions:
\begin{equation}
\label{eq:twoparit}
\phi(y) \rightarrow  \eta \,\phi(\pi R/2 - y) \quad,\quad
\phi(y) \rightarrow  \eta' \,\phi(\pi R/2 - y) \;.
\end{equation}

The point is that both symmetries defined in the above equation
are broken by the boundary conditions: 
there are fields with different boundary conditions at $0$ and $\pi R/2$ and this is not 
compatible with the reflection that exchanges the two fixed points. 
The same may happen also on $S^1/\Z_2$ if we put fixed-point interactions that are different at 
$0$ and $\pi R$, so that the two fixed points are physically distinguishable. The simple reason why
on $S^1/(\Z_2 \times \Z^\prime_2)$ we cannot find a global 5d parity symmetry is that we have
already gauged the most general discrete symmetry group on $S^1$~\cite{Barbieri:2001dm}.
However, 
even if it is impossible to define a global symmetry to avoid parity-odd operators, if we do 
not introduce odd operators from the beginning they will not be radiatively generated. 
This is because only the boundaries break the symmetries  of eq.~(\ref{eq:twoparit}),
 so that only non-local terms will be influenced
\footnote{We are in a situation similar to a particle physics experiment 
held in a laboratory that is not parity symmetric: we do not expect that this will induce 
local interactions that violate parity!}. 
To put this idea on more solid grounds we have to define this new symmetry locally. This is just a diffeomorphism
that acts like  one of the reflections (\ref{eq:twoparit}) in an open set inside the fundamental 
domain and trivially outside. Obviously we need more than one chart to define this diffeomorphism over the full
space. The boundaries break global parity very much in the same 
way as they break translations.

The presence of this local parity invariance ensures that odd terms will not be radiatively induced starting from a theory 
with only even terms.  
Note that, as this parity cannot be defined globally, it cannot be used to classify the physical states. Similarly
changing the sign of the coefficients of all the odd operators in the Lagrangian leads to inequivalent theories.
A relevant example, which we will encounter, showing this inequivalence is a Dirac fermion mass.

Although the introduction of odd operators is fully consistent, something singular has to happen at the orbifold fixed
points: either the coefficients of the odd operators have to change sign or the vacuum expectation value of an odd   
scalar has to jump. In the absence of this singular behavior the physics on the orbifold is the ``smooth'' projection of
the physics on the full circle: in the presence of odd operators the relation with the original $S^1$ theory may be
subtle, as is shown in the calculation of anomalies through index theorems in appendix \ref{fuji}.

\section{Anomalies}
\label{pv}

In this section we want to study the simple case of 5d spinor electrodynamics
compactified on an $S^1/(\Z_2\times \Z_2^\prime)$ orbifold.
Before writing the Lagrangian and the fields, let us briefly establish our notation.
We label our full space-time directions by Latin capital
letters $M,N,\dots=0,1,2,3,5$, where 5 indicates the compact one, and we indicate
the non-compact directions by the Greek letters $\mu, \nu,\dots=0,1,2,3$.
We use the time-like Lorentz metric $\eta_{MN}={\rm diag}(+,-,-,-,-)$.
The 5d Dirac matrices $\Gamma^M$ are
\begin{equation}
\Gamma^{\mu}=
\left (\begin{array}{cc} 0 & \sigma_+^{\mu} \\ \sigma_-^{\mu} & 0 \end{array}\right )
 \equiv \gamma^\mu \;,\quad
\Gamma^5=-i\left (\begin{array}{cc} -1 & 0\\ 0 & 1 \end{array} \right ) \equiv -i\gamma^5
\;,\quad 
\sigma_{\pm}^{\mu}=(1,\pm \sigma^i) \;,
\end{equation}
with $\sigma^i$ the Pauli matrices and $\gamma^\mu$, $\gamma^5$ the usual 4d matrices.
We associate to the fields the labels $(++)$, $(+-)$, $(-,+)$ and $(-,-)$,
depending on their parity under the $\Z_2\times \Z_2^\prime$ transformations.
We are interested in the case where
$A_\mu=(+,+)$ and $A_5=(-,-)$ while the charged fermion satisfies
$\psi(-y)=\gamma^5\psi(y)$ and $\psi(\pi R - y)=-\gamma^5\psi(y)$.
Then if we write $\psi$ in terms of bispinors
$\psi=(\chi,\bar\chi_c)$ we have $\chi=(-,+)$ and $\chi_c=(+,-)$. The Lagrangian is
\begin{equation}
{\cal L} = \int_0^{2\pi R}\!\!\! dy\, \left[-\frac{1}{4g^2} F^{MN}F_{MN}+\bar \psi\,
i\Gamma^{M} \big(\partial_M -i A_M\big) \psi \right] \;.
\end{equation}
Even though we have not included a 5d mass term, no massless fermion
survives the compactification
because of the $\Z_2\times \Z_2^\prime$ parity assignments.
The Kaluza--Klein decomposition of $\psi$ is
\begin{gather}
\label{decompPsi}
\chi(x,y)=\sum_{n\geq 0} \xi^{-+}_{n}(y) \chi^{n}(x)\quad,\quad
\bar \chi_c(x,y)=\sum_{n\geq 0} \xi^{+-}_{n}(y) \bar\chi_c^{n}(x) \;,\\
\label{masslesswf}
\xi^{+-}_{n}(y) = a_{2n+1} \cos \frac{2n+1}R y \quad,\quad
\xi^{-+}_{n}(y) = a_{2n+1} \sin \frac{2n+1}R y \;,
\end{gather}
where $a_{k}=1/\sqrt{\pi R}$ for $k\not =0$ and $a_{0}=1/\sqrt{2\pi R}$.
These represent at each level $n$ a Dirac fermion with mass $M_n = (2n+1)/R$.
The gauge field instead decomposes as
\begin{gather}
\label{decompA}
A_\mu(x,y)=\sum_{n\geq 0} \zeta^+_{n}(y) A_\mu^{n}(x)\quad,\quad
A_5(x,y)=\sum_{n\geq 1} \zeta^-_{n}(y) A_5^{n}(x) \;,\\
\zeta^+_{n}(y) = a_{2n} \cos\frac{2n}{R}y \quad,\quad
\zeta^-_{n}(y) = a_{2n} \sin\frac{2n}{R}y \;,
\end{gather}
which represents at each level a massive vector with mass $2n/R$.
Notice that the orbifold projections act differently
on the two chiralities so that this is a 4d chiral gauge theory. However 4d parity $P$
is still preserved combined with
a 5d reflection about the midpoint $y=\pi R/4$ of the fundamental domain:
$y\to\pi R/2-y$. The action of $P$ on the fields is
\begin{equation}
\begin{split}
 \chi(x_0,x_i,y) &\to \bar\chi_c(x_0,-x_i,\pi R/2-y), \\
 \bar\chi_c(x_0,x_i,y) &\to \chi(x_0,-x_i,\pi R/2-y),
\end{split} \quad
\begin{split}
 A_\mu(x_0,x_i,y) &\to A^\mu(x_0,-x_i,\pi R/2-y), \\
 A_5(x_0,x_i,y)&\to -A_5(x_0,-x_i,\pi R/2-y)\;.
\end{split}
\end{equation}
This transformation leaves the boundary conditions unaffected, and thus
$P$ survives after compactification.
The KK gauge bosons $A_\mu^{n}$ have parity $(-)^{n+1}$,
so that for $n$ even (odd) they couple through a vector (axial)
current; also the scalars $A_5^{n}$ have parity $(-)^n$ and
couple through a scalar (pseudoscalar) density for $n$ even (odd).
In the same way it is easy to see that the system respects
a charge conjugation $C$, defined as
\begin{equation}
\label{chargeC}
\begin{split}
 \chi(y) &\to \chi_c(\pi R/2-y), \\
 \chi_c(y) &\to \chi(\pi R/2-y),
\end{split} \quad \quad
\begin{split}
 A_\mu(y) &\to -A_\mu(\pi R/2-y), \\
 A_5(y) &\to A_5(\pi R/2-y)\;.
\end{split}
\end{equation}
The KK modes transform under $C$ as $A_\mu^{n}\to (-)^{n+1}A_\mu^{n}$ and
$A_5^{n}\to (-)^{n+1} A_5^{n}$.
The use of these discrete symmetries will be useful in the following
discussion on the anomaly.

It is known that on Minkowski spacetimes of odd dimensionality
Abelian gauge theories are never anomalous
(there are however global anomalies
in the non-Abelian case \cite{redlich,alvarez}). However in the case at hand
there are boundaries and one should be careful.
In fact, as we already remarked,
the system we are considering represents a chiral gauge theory from the 4d view-point:
there are both vector ($V$) and axial ($A$) massive gauge fields.
Very much as it happens in 4d, there are in principle UV ambiguities in the
definition of the three-point functions of the type $AVV$ and $AAA$.
Moreover, as in 4d, one should not worry about $VVV$ and $AAV$, which
vanish by charge conjugation. As noticed in ref.~\cite{Arkani-Hamed:2001is} it is
useful to arrange the fermions in an infinite vector of Dirac spinors
$\Psi=(\psi_1(x),\psi_2(x),\dots)$, $\psi_n(x)=(\chi^{n}(x),\bar\chi^{n}_c(x))$.
The gauge field is just a (infinite) matrix acting on this vector
\begin{equation}
{\cal V}_\mu(x)+\gamma^5 {\cal A}_\mu(x)=\int\! dy\, A_\mu(x,y)\, (Q_V(y)+\gamma^5 Q_A(y)) \;,
\end{equation}
where the charge matrices $Q_V$ and $Q_A$ are determined by the fermion wave functions
\begin{equation}
\left(Q_{V,A}(y)\right)_{mn} = \frac{1}{2}
 \left[\xi^{+-}_{m}(y)\xi^{+-}_{n}(y) \pm \xi^{-+}_{m}(y)\xi^{-+}_{n}(y) \right] \;.
\end{equation}
Similarly we can write the coupling to $A_5$ via a scalar charge matrix
\begin{equation}
\begin{gathered}
{\cal A}_5(x) =\int\! dy\ A_5(x,y)\, \Omega(y) \;,\qquad
 \Omega(y)=Q_{S}(y)+\gamma^5 Q_{P}(y) \;,\\ \left(Q_{S,P}(y)\right)_{mn} = \frac{1}{2}
 \left[\xi^{-+}_{m}(y)\xi^{+-}_{n}(y) \mp \xi^{+-}_{m}(y)\xi^{-+}_{n}(y) \right]\;.
\end{gathered}
\end{equation}
so that the fermion Lagrangian is
\begin{equation}
{\cal L} = \bar\Psi\gamma^\mu\left(i\partial_\mu +
{\cal V}_\mu(x)+\gamma^5 {\cal A}_\mu(x)\right)\Psi -
\bar\Psi \left( {\cal M}+i{\cal A}_5(x)\right )\Psi \;,
\label{messyL}
\end{equation}
where $\cal M$ is the mass matrix: ${\cal M}_{mn} = \delta_{mn} M_n$.
The gauge transformations are represented~by
\begin{equation}
\begin{aligned}
\delta\left[{\cal V}_\mu(x)+\gamma^5 {\cal A}_\mu(x)\right] &=
 \int\! dy\ \partial_\mu a(x,y)\, (Q_V(y)+\gamma^5 Q_A(y)) \\
\delta {\cal A}_5(x) 
&=\; \int\! dy\ \partial_y a(x,y) \, \Omega(y) \\
\Psi(x) &\to \exp \left[i\!\int\! dy\ a(x,y)\, (Q_V(y)+\gamma^5 Q_A(y))\right]\, \Psi(x)
\end{aligned}
\end{equation}
under which invariance of the Lagrangian (\ref{messyL}) follows thanks
to the relation
\begin{equation}
\partial_y\Omega(y)=[{\cal M},Q_V(y)]+\gamma^5\{ {\cal M},Q_A(y)\} \;,
\end{equation}
which is nothing but the statement that ${\cal M}\sim \gamma^5\partial_5$ on
the fermion space.
In the same way, the 5d divergence of the fermion current $\partial_MJ^M$
at tree level is simply given by the variation
of the Lagrangian (\ref{messyL}) under a gauge rotation of the fermions,
keeping the gauge fields unchanged
\begin{equation}
\partial_MJ^M=
\partial_\mu \bar\Psi\gamma^\mu (Q_V+\gamma^5 Q_A)\Psi -i\,
\bar\Psi [{\cal M},Q_V]+\gamma^5\{ {\cal M},Q_A \}\Psi \;.
\label{5divergence}
\end{equation}
By the last equation, anomalies of the 5d current can be studied
by using the known 4d results \cite{Arkani-Hamed:2001is}. Indeed, eq.~(\ref{5divergence}) is simply
a generalization of the familiar expression
$\delta {\cal L}=\partial_\mu J_A^\mu-2im\, \bar\psi\gamma^5\psi$ encountered in
the study of the chiral anomaly.
By applying the known 4d results it is then straightforward to
calculate the anomaly in the current conservation
\begin{equation}
\partial_MJ^M(x,y)=\frac{1}{96\pi^2} \left[ {\rm Tr}
 \left (Q_L(y){\cal F}_L(x)\tilde{\cal F}_L(x)\right ) -
 {\rm Tr}\left (Q_R(y){\cal F}_R(x)\tilde{\cal F}_R(x)\right ) \right] \; ,
\label{4danomaly}
\end{equation}
where $Q_{R,L}=Q_V \pm Q_A$ and ${\cal F}_{R,L}={\cal F}_V \pm {\cal F}_A$.
Using the completeness properties of the fermion eigenmodes, this expression
can be written in a very simple and local form \cite{Arkani-Hamed:2001is}.
For the model at hand, it was calculated in ref.~\cite{Scrucca:2001eb},
finding~\footnote{Actually our result is 1/3 that of ref.~\cite{Scrucca:2001eb} as
we must symmetrize with respect to the external vectors. 
See also appendices \ref{regulators} and \ref{fuji}.}
\begin{equation}
\partial_M J^M(x,y)= \frac{1}{4}
\left[\delta(y)-\delta(y-\pi R/2)+\delta(y-\pi R)-\delta(y-3\pi R/2)\right]
\, \frac{-1}{96\pi^2} F_{\mu\nu}\tilde F^{\mu\nu} \;.
\label{anomaly}
\end{equation}
The form of this result, with ``equal and opposite anomalies'' at the two boundaries,
could have easily been anticipated by simple arguments.
First, the anomaly, if it exists, should be localized at the boundaries:
inside the bulk, the UV properties of the triple current correlator
are insensitive to the presence of the boundaries, then current conservation follows from
the absence of anomalies in five dimensions. Second, in our case there is a 4d parity
symmetry under which the two boundaries are exchanged, and for which the gauge parameter
$a(x,y)$ is a scalar.
Then, as $F\tilde F$ is parity-odd, the contributions at the two boundaries should be
equal and opposite.

The result (\ref{anomaly}) implicitly assumes the use of a mode-by-mode regularization procedure 
that breaks gauge invariance explicitly. The same result can also be reproduced from a 5d point 
of view, for example by using a higher-derivative deformation of the fermionic kinetic term,
as shown in appendix \ref{regulators}. Alternatively, one can also adopt Fujikawa's approach,
in which the anomaly is associated to a Jacobian in the measure of the functional integral
defining the effective action; see appendix \ref{fuji}.

\subsection{Chern--Simons term}

Let us keep arguing from a 4d perspective. It is well-known that the 4d anomaly is
determined up to the addition of local counterterms of the type
$A^{m}_\mu A^{n}_\nu\partial_\rho A^{r}_\sigma\epsilon^{\mu\nu\rho\sigma}$.
When at least two of the vectors are different,
these counterterms can be used to shift the anomaly on just one of the three vertices,
satisfying current conservation on the other two.
This is what happens for the chiral anomaly, where the anomaly can be shifted to the
axial vector current preserving vector current conservation. Now, eq.~(\ref{4danomaly}) corresponds to
symmetrizing the anomaly with respect to the three (in general different) external legs and so, given
a regulator, it represents a particular choice of local counterterms. 
In general, eq.~(\ref{4danomaly}) should be considered up to the addition of such local counterterms. 
Indeed in our case there is yet another
class of counterterms that can be added in order to shift (or eliminate) the anomaly. Of our big gauge group,
only one $U(1)$ factor, associated to the zero mode $A_\mu^{0}$, corresponds to a linearly realized symmetry. The KK
modes correspond to a Higgsed gauged symmetry for which ${A}_5^{n}$ are the (eaten) Goldstone bosons. Then one can
also add terms of the type $A_5^{m}F_{\mu\nu}^{n} \tilde {F}^{r\,\mu\nu}$, which can affect eq.~(\ref{anomaly})
since $A_5$ transforms inhomogeneously. In looking at this very special 4d system, one may then wonder
whether the anomaly can be eliminated by a specific choice of these two classes of counterterms.
Moreover, having the 5d picture dear, one may also ask whether a choice of
counterterms exists which builds up
into a local 5d operator. The answer to these questions is easy.
Thinking directly in 5d, one immediately realizes that the addition of the Chern--Simons (CS)
term
\begin{equation}
{\cal L}_{CS}=\frac{1}{192\,\pi^2}\int_0^{\pi R/2}\!\!\! dy\
\epsilon^{MNOPQ}\, A_M F_{NO} F_{PQ}
\end{equation}
exactly cancels the anomaly of eq.~(\ref{anomaly}).
It is easy to see that, when decomposed in KK
modes, the CS term corresponds to local operators belonging to the two
classes we mentioned above.

Note that in the above equation we have knowingly written the CS term by working
on the single covering of the orbifold. When working on the full circle $S^1$, the CS would be
\begin{equation}
{\cal L}_{CS}=\frac{1}{4}\; \frac{1}{192\pi^2} \int_{S^1}\!\! dy\
\eta(y)\,\epsilon^{MNOPQ}\, A_M F_{NO} F_{PQ} \;,
\label{cscircle}
\end{equation}
where 
\bea
\eta(y)= \left\{\begin{array}{l}
+1\;,\;\; y \in \,\big(0,\pi R/2\big)\, \cup \,\big(\pi R,3\pi R /2\big) \medskip\ \\
-1\;,\;\; y \in \,\big(\pi R/2,\pi R\big)\, \cup \,\big(3\pi R /2, 2 \pi R\big) 
\end{array}\right. \;.
\eea
The almost obvious fact that such a CS term with a jumping coefficient could be used
to cancel the anomaly was also noticed in ref.~\cite{Scrucca:2001eb}, but not fully exploited.
That was also because of some skepticism about the consistency of this parity-odd term  
with the orbifold projection.
Indeed the CS term is odd, but only under the local 5d parity defined in section \ref{odd}.
As we explained there, the Lagrangian in eq.~(\ref{cscircle}) actually is perfectly 
consistent with the discrete symmetry that has been gauged to implement the orbifold 
projection.
Of course, if we defined our field theory just by working on the segment and 
by assigning to fields either Dirichlet ($+$ fields) or Neumann ($-$ fields) conditions
at each boundary, discontinuous parameters would not arise and there would be no orbifold
symmetry to be confused about. Then there would manifestly be no issue.

The whole discussion can be easily generalized to a non-abelian gauge group.
It is well-known that non-abelian global anomalies may exist, even in 5d Minkowski space,
in a theory with an odd number of fermions~\cite{redlich}.
In this case the action changes by $\pm \pi n$ under a homotopically non-trivial gauge transformation with
winding number $n$; to restore gauge invariance a Chern--Simons term has to be added, violating in this 
way the 5d parity. 
On an orbifold, because of the presence of boundaries, the situation is quite different, as the anomaly is not
a global effect but arises locally at the fixed points. For example, a theory with two fermions in
5d Minkowski space is not anomalous irrespectively of their gauge representation, while on the orbifold
this is true only if the fermions are in conjugate representations.

\subsection{Gauge-invariant regularization}

We went through all the above discussion of the anomaly because we wanted to better interpret the
result of ref.~\cite{Scrucca:2001eb}. Instead we could have started by directly looking for a gauge-invariant 
regulator. Finding one would imply that there are no anomalies. In fact, this model can be 
regulated \`a la Pauli--Villars (PV). First notice that, compatibly with gauge invariance and the orbifold 
boundary conditions, we can add a ``jumping'' fermion mass
\begin{equation}
{\cal L}_{\rm mass}=-\int_{S^1}\!\! dy\ m\, \eta(y)\ \bar\psi\psi \;.
\end{equation}
Similarly, we can then add a PV Dirac spinor $\Omega$ with wrong statistic and
jumping mass $\eta(y) M$.
Secondly we must study the PV spectrum when $|M|\to \infty$ to make sure that all the modes
become infinitely heavy. We recall that this condition is not guaranteed in the presence of
a jumping profile, as localized massless states can arise. For instance, on
$S^1/\Z_2$ with one Dirac fermion, there is one massless chiral mode, which must survive the 
addition of a jumping mass $m\,c(y)$ for continuity reasons \cite{Georgi:2000wb}.
Since parity $P$ is a symmetry also in the presence of the mass
term, the fermion KK modes will keep forming Dirac fermions.
In order to study the spectrum, we need to consider only one chirality, say 
$\Omega_L$, which is a $(-+)$ field. The eigenvalue equation is
\begin{equation}
(-\partial_y+M\eta(y))(\partial_y+M\eta(y))\, \xi^{-+}_{n}=M_n^2 \, \xi^{-+}_n
\label{prop}
\end{equation}
and the eigenvalues are given by the solutions of the equation
\begin{equation}
-\frac{\lambda_n}{M}= \tan \frac{\lambda_n\pi R}{2}
\label{implicit}
\end{equation}
through the relation
\begin{equation}
M_n^2=M^2+\lambda_n^2 \;.
\label{eigenvalue}
\end{equation}
The behavior of the spectrum for $M\to \infty$ depends on the
sign of $M$. For $M$ positive, all the solutions $\lambda_n$ of eq.~(\ref{eigenvalue})
are real, and then each $M_{n}\to \infty$ when $M\to +\infty$.
For  $M$ negative, and more precisely for $M<-2/(\pi R)$,
 there is also an imaginary solution $\lambda_0=i\tilde \lambda$.
As $M\to -\infty$, $\lambda_0\to iM$ and
the corresponding eigenvalue $M_0\to 0$  (one has $M_0\sim {\sqrt 2} M e^{-M\pi R/2}$).
This asymptotically massless mode is localized at $y= \pi R/2$,
while its right-handed partner is localized at the other boundary.
This result is fairly intuitive. For $|M|\gg 1/R$, positive or negative, the presence
of localized light fermions at each boundary can be established by disregarding
the effects of the other boundary.
Only for $M$ negative are there localized zero modes compatible with the $\Z_2$ or
$\Z_2^\prime$ boundary condition at each fixed point.
A (set of) PV field(s) with positive mass is then our candidate regulator. As we will see in a moment,
the selection of a sign for the regulator mass $M$ is associated to the sign of the CS term which
cancels the anomaly.

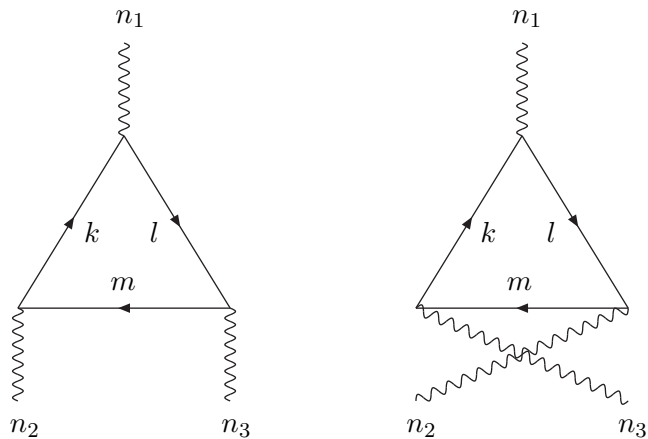
\begin{figure}[h]
\begin{center} 
\begin{picture}(250,170)(0,-30)
\put(47,128){$n_1$}
\put(7,-27){$n_2$}
\put(87,-27){$n_3$}
\put(35,45){$k$}
\put(60,45){$l$}
\put(45,27){$m$}
\put(197,128){$n_1$}
\put(157,-27){$n_2$}
\put(237,-27){$n_3$}
\put(185,45){$k$}
\put(210,45){$l$}
\put(195,27){$m$}
\ArrowLine(90,20)(10,20)
\ArrowLine(10,20)(50,85)
\ArrowLine(50,85)(90,20)
\Photon(10,20)(10,-15){2}{7}
\Photon(50,85)(50,120){2}{7}
\Photon(90,20)(90,-15){2}{7}
\ArrowLine(240,20)(160,20)
\ArrowLine(160,20)(200,85)
\ArrowLine(200,85)(240,20)
\Photon(160,20)(240,-15){2}{14}
\Photon(200,85)(200,120){2}{7}
\Photon(240,20)(160,-15){2}{14}
\end{picture}
\caption{\label{diag} The two diagrams contributing to the vector three-point function.}
\vskip -10pt
\end{center}
\end{figure}

We have explicitly checked that the fermion loop short distance singularities are regulated to an
arbitrary degree by adding a combination of PV fields. Consider for instance the vector 3-point function.
It is convenient to work in Euclidean momentum space along the compact directions and in position space
for the fifth direction. We can for instance calculate the fermion propagator in the limit in which
we take the second fixed point to infinity; for $y,y^\prime >0$ we have
($\tilde p \equiv \sqrt{p^2 + m^2}$):
\begin{equation}
\begin{split}
S(p;y,y^\prime) = \frac1{2 \tilde p} & \Big[\frac{e^{- \tilde p (y+y^\prime)}}{\tilde
p + M} \big(\tilde p \,\gamma^5 - \not\!p - M\big)
\big(\tilde p \,\gamma^5 +M\big) \\ & + e^{-\tilde p |y-y^\prime|} \big(-\tilde p \,
\gamma^5\cdot{\rm sgn}(y-y^\prime) + \not\! p +M \big)\Big] \;,
\end{split}
\end{equation}
with obvious extension to $y,y^\prime <0$.
In this mixed notation the 3-point vertex depends on both the three 4d external momenta $(p_1,p_2,p_3)$
and the three  positions in the fifth dimension $(y_1,y_2,y_3)$, and of course there is a $d^4p$
integral over the virtual 4d momentum. UV ambiguities arise only when the $y_i$ coalesce to a point; otherwise,
the loop momentum integral converges exponentially thanks to the propagator terms 
$\exp(-\tilde p|\Delta y|)$.
However, for coinciding $y$'s, the integrand is just an {\it analytic} function of the momenta and masses.
So it can be made to vanish for $p\to \infty$ like $1/p^{k}$ with arbitrary $k$ by the addition of a suitable number
of PV fields with masses $M_i$ and charges $q_i$ satisfying relations of the type
\begin{equation}
\sum_i q_i^3=1 \; , \quad\quad \sum_i q_i^3 M_i^n=0 \quad {\rm for}\ n>0 \; ;
\end{equation}
here the first relation ensures that the leading divergence associated with the physical fermion
(with charge $q=1$) is canceled. Then all the possible UV ambiguities in the gauge boson vertices
can be eliminated in a manifestly gauge-invariant way and we conclude that the theory is not anomalous.
The presence of a gauge-invariant regulator allows to power-count the UV divergences
by using gauge invariance. For 3-point vertices it is straightforward to see that the result must be finite:
the only bulk operator that could, in principle, be logarithmically divergent is the CS term; however the CS cannot be
made fully gauge invariant by the addition of local terms at the boundaries. Still, in order to properly define
the 3-point vertex the addition of at least one PV field is necessary. In order to see this, it is easier to
work with KK-modes rather that with the mixed propagator. 
The basic story is the following. The only vertices that are potentially anomalous and need regulation are those 
for which the 3 vector wave numbers satisfy the relation $n_1+n_2+n_3={\rm odd}$ \cite{Scrucca:2001eb}. 
(These are the vertices of the type $AVV$ and $AAA$. The $VVV$ and $AAV$ vanish by charge conjugation, while 
those involving one external scalar are finite.)
For these diagrams, there is only a finite number of choices of the internal fermion fifth momentum: there is
no infinite KK sum and the situation is therefore exactly the same as in 4d, with a linear divergence. 
This is essentially due to the fact that $n_1+n_2+n_3={\rm odd}$ violates momentum conservation in the fifth 
direction. The presence of these diagrams is just a consequence of the boundary. 
Given the external bosons, there are two such diagrams (see fig.~\ref{diag}).
They are individually linearly divergent, but although their sum is finite the result is ambiguous since
it depends on the routing of the integration momentum on each diagram. The addition of one PV regulator field
resolves this ambiguity while preserving gauge invariance. Now for each diagram of fig.~\ref{diag} we subtract an
(infinite) series of PV diagrams. After the subtraction the linear divergence is eliminated and there is
no longer a momentum routing ambiguity. The series of PV diagrams works like a generalization of multi-PV
regulators. What matters is that the leading divergence of the sum of PV matches exactly the
leading divergence of the physical diagram. This is explained in detail in appendix \ref{regulators}.

We have argued that the anomaly can be eliminated either by adding a CS with jumping coefficient or by regulating the
theory via a PV with jumping mass. What is the relation between the two approaches?
The answer is pretty simple. Notice first of all that both the CS and the fermion mass are odd
under the (local) 5d parity.
Then, by symmetry and power counting, it 
is expected that for $M\to\infty$ there should remain a finite
CS term. We can easily calculate its coefficient by looking at the triple
vertex for three points inside the bulk. 
The term we are interested in corresponds to a single mass insertion and,
for $M\to \infty$, it does not feel the
presence of the boundaries, so that the result is the same one obtains in an infinite fifth dimension. 
By explicit calculation we find  that the effective CS term is
\begin{equation}
\label{eq:CSPV}
{\cal L}_{CS}^{eff}=\frac{1}{4}\; \frac{1}{192\,\pi^2}\frac{M}{|M|}
\int_{S^1}\!\! dy\ \eta(y)\, \epsilon^{MNOPQ}\, A_M F_{NO} F_{PQ} \;,
\end{equation}
which is precisely what we need to cancel the anomaly provided $M$ is positive.
But $M$ positive is indeed
the requirement imposed on our regulator by the decoupling requirement.
Therefore the two pictures nicely match.

\section{\label{FI} The Fayet--Iliopoulos term}

Given the connection between FI terms and mixed $U(1)$-gravitational anomalies in 4d, it is 
natural to ask whether a similar relation persists in 5d orbifold theories as well.

It is well-known that 4d theories with rigid supersymmetry admit 
a FI term for $U(1)$ vector multiplets: ${\cal L}_{FI} = [\xi V]_D$.
This term is manifestly supersymmetric. It is also invariant under the super-gauge transformation 
$V \rightarrow V + \Lambda + \bar \Lambda$, since $[\Lambda]_D = [\bar \Lambda]_D = 0$. 
However, its presence destabilizes the vacuum and triggers the spontaneous breaking 
either of supersymmetry or of the $U(1)$ gauge symmetry (or of both). 

In 4d supergravity theories, the situation is different: the naive extension of the 
rigid FI term is not gauge-invariant by itself. In the superconformal 
approach \cite{superconformal}, with a compensator chiral multiplet $S_0$, a rigid FI term would be promoted to 
$[S_0 \bar S_0 \xi V]_D$, which is no longer invariant under 
$V \rightarrow V + \Lambda + \bar \Lambda$ since $[S_0 \bar S_0 \Lambda]_D$ and
$[S_0 \bar S_0 \bar \Lambda]_D$ are non-zero. To write a gauge-invariant generalization
of a FI term, $S_0$ must transform under the $U(1)$ symmetry: a term of the form
$[S_0 \bar S_0 e^{\xi V}]_D$ will be invariant, provided the compensator undergoes a 
super-Weyl transformation $S_0 \rightarrow e^{-\xi \Lambda} S_0$, 
$\bar S_0 \rightarrow e^{-\xi \bar \Lambda} \bar S_0$ \cite{Barbieri:1982ac,Ferrara:1983dh}. 
This implies that the $U(1)$ symmetry must in fact be a gauged version of the $U(1)_R$ 
symmetry: the FI term will induce a non-vanishing charge for the gravitino. 
In component fields we can check that the linear term $[S_0 \bar S_0 \xi V]_D = \xi e (D 
+ \frac i4 \epsilon^{\mu \nu \rho \sigma} \bar \psi_\mu \gamma_\nu \psi_\rho A_\sigma
+ \cdots )$ indeed contains a coupling of the $U(1)$ vector field $A_\mu$ to the gravitino $\psi_\mu$,
beside the rigid FI term.

An alternative way to have a FI-like term in local supersymmetry is to introduce the superfield $V$ 
in combination with a chiral multiplet $\Phi$ transforming non-homogeneously under the 
$U(1)$: $\Phi \rightarrow \Phi - \Lambda$. One can then write a generic gauge-invariant 
K\"ahler potential of the form $K(\Phi + \bar \Phi + V)$; the linear piece in the expansion
of $K$ will give rise to a FI term $K^\prime(\langle\Phi+\bar\Phi\rangle) V$
\cite{Dine:xk}. In general, this mechanism gives, however, a mass term
$\frac 12 K^{\prime\prime}(\langle\Phi+\bar\Phi\rangle) V^2$ for the 
$U(1)$ gauge boson.

Summarizing, in four dimensions, a FI term is in general not gauge-invariant in the 
presence of gravity. If the theory is affected by mixed $U(1)$-gravitational anomalies 
this term is radiatively generated. More precisely, the anomaly 
and the FI term will be simultaneously generated at one loop, with a common 
coefficient given by the trace of the $U(1)$ charge over the spectrum of the model. 

Given its connection with mixed anomalies in the 4d supergravity, we are interested in studying 
the possibility to have FI terms at the orbifold fixed points in the 5d theory. We will see
that the situation is quite different with respect to the 4d case.
Indeed, from the 4d point of view, the chiral multiplet $\Phi$ containing 
$A_5$ transforms inhomogeneously under gauge transformations and it has a 
K\"ahler potential of the form $K(\Phi + \bar\Phi - \partial_5 V)$. As we shall see, this 
will allow for a consistent FI term, whose gauge variation is canceled by bulk terms. 
Furthermore, such a term will spontaneously break neither supersymmetry nor gauge 
symmetry, contrarily to what happens in 4d. Moreover we will see that the radiative generation
of FI operators at the orbifold fixed points is not forbidden even in the absence of mixed anomalies.

\subsection{\label{z2} The $S^1/\Z_2$ model}

We begin by considering the $S^1/\Z_2$ case, where one of the supersymmetries remains 
unbroken and one can efficiently use the superfield formalism with respect to this $N=1$
\cite{Arkani-Hamed:2001tb}. An $N=2$ vector multiplet decomposes into an $N=1$ vector multiplet 
$V=(A_\mu,\lambda_1,D)$ plus a chiral multiplet $\Phi = (\frac 1{\sqrt{2}}(\Sigma + i A_5), 
\lambda_2,G)$, transforming in the adjoint representation of the gauge group 
\footnote{In terms of $N=2$ auxiliary fields, one has: $D=X_3 - \partial_5 \Sigma$
and $G = X_1 + i X_2$ \cite{Mirabelli:1997aj}.}. 
Similarly, an $N=2$ hypermultiplet decomposes into two chiral multiplets 
$H=(\phi,\chi,F)$ and $H_c=(\phi_c,\chi_c,F_c)$. The kinetic Lagrangian of a hypermultiplet 
interacting with a $U(1)$ vector multiplet, in the presence of a FI term is 
\begin{equation}
\begin{split}
{\cal L}_K = \;&\frac 1{g^2}\int d^4 \theta \Big[2\,\xi(y)\, V 
+ (\partial_5 V - \frac 1{\sqrt{2}}(\Phi + \bar \Phi))^2 \Big]
+ \frac 1{4g^2} \int \Big[d^2\theta \, W^\alpha W_\alpha + {\rm h.c.}\Big] \label{LKS} \\
& + \int \! d^4\theta \, \Big[\bar H e^{qV} H + H_c e^{-q V} \bar H_c \Big]
- \int \Big[d^2\theta \, H_c \big(\partial_5 - \frac q{\sqrt{2}} \Phi - m(y)\big) H 
+ {\rm h.c.}\Big] \;,
\end{split}
\end{equation}
where $g$ is the gauge coupling and $q$ is a dimensionless charge.
Notice the appearance of the covariant derivative in the extra dimension $\partial_5 - 
\frac q{\sqrt{2}} \Phi$ \cite{Arkani-Hamed:2001tb,Hebecker:2001ke}. The above Lagrangian is 
manifestly supersymmetric and the orbifold projection implies $m(-y) = - m(y)$, and $\xi(-y) = 
\xi(y)$. Furthermore, $N=2$ bulk supersymmetry would imply that $m(y)$ and $\xi(y)$ are piecewise constant: 
$m(y) = m \cdot c(y)$ and $\xi(y)=\xi_0$ plus contributions localized at the orbifold fixed points.
For the bosonic components, we find (after eliminating the auxiliary fields $G$, $F$ 
and $F_c$):
\begin{equation}
\begin{split}
{\cal L}_K =\;& \frac 1{g^2}\Big[-\frac 14 \, F_{MN} F^{MN} + \frac 12 \,\partial_M \Sigma\, \partial^M \Sigma 
+ \frac 12 (D + \partial_5 \Sigma)^2 - \xi(y) D \Big] \\
& +\, |{\cal D}_M \phi|^2 - \Big[(m(y) + q \Sigma)^2 + \partial_5 m(y)
+ q \big(D + \partial_5 \Sigma\big) \Big] |\phi|^2 \\
& +\, |{\cal D}_M \phi_c|^2 - \Big[(m(y) + q \Sigma)^2 - \partial_5 m(y)
- q \big(D + \partial_5 \Sigma\big)\Big] |\phi_c|^2 \;.
\label{LKC}
\end{split}
\end{equation}
Notice the appearance of the $\partial_5 m$ mass terms. In the relevant case of a piecewise constant $m$,
these contributions are localized at the boundary. Their presence ensures that the 5d-supercurrent $S_M^\alpha$
is conserved everywhere, a necessary condition to define the supergravity version of the model.
The equation of motion for $D$ is
\be
D = \xi(y) - \partial_5 \Sigma + q\,g^2\, (|\phi|^2 - |\phi_{c}|^2) \;.
\label{Dflat}
\end{equation}
We see that differently from the 4d case, the neutral scalar $\Sigma$ appears in the equation
of motion for $D$. This is due to the non-homogeneous transformation law of the chiral multiplet
$\Phi$ and allows us to find a supersymmetric ($D=0$) and gauge-invariant
vacuum state even in the presence of the FI term.
In fact, the $D$-flatness condition can be solved with $\langle\phi\rangle = \langle\phi_c\rangle = 0$ 
and $\partial_5\langle\Sigma(y)\rangle = \xi(y)$. For generic FI terms at
the two fixed points it is in general impossible to satisfy the condition $D=0$ at both
boundaries. An ``integrability condition'' $\int \! dy \, \xi(y) = 0$ is necessary to 
have a supersymmetric vacuum state: we must have opposite FI terms at the two sides. 
This requirement is equivalent to having a vanishing
FI term in the effective 4d action: this was expected, because, as we saw, a 4d FI term is not
compatible with a gauge-invariant supersymmetric vacuum.
If the integrability condition is satisfied, the scalar potential is
\be
\label{eq:Vscal}
V_{\rm scalar} = \Big[(m (y) + q \Sigma)^2 (|\phi|^2 + |\phi_{c}|^2) 
+ \partial_5(m(y) + q \Sigma) (|\phi|^2 - |\phi_{c}|^2) \Big] \;.
\end{equation}
The net effect of the FI terms is then to shift the mass terms by the VEV of $\Sigma(y)$.

A FI term satisfying the integrability condition is in fact equivalent to shifting $\Phi$ by a 
stepwise constant. This can be easily seen from the Lagrangian (\ref{LKS}): a non-vanishing 
$\langle \Phi + \bar \Phi\rangle = \sqrt{2} \langle \Sigma(y) \rangle$ in the second term 
generates a term $-2 \langle \Sigma(y) \rangle [\partial_5 V - \frac1{\sqrt{2}}
(\Phi + \bar\Phi)]_D$, which, in rigid supersymmetry, is equivalent to a FI term with $\xi(y) 
= \partial_5 \langle \Sigma(y) \rangle$ after integration by parts. 

This relationship between the FI term and bulk operators turns out to be important to understand the 
supergravity extension. In particular it says that boundary  FI terms are equivalent to the bulk operator 
$c(y)[\partial_5 V - \frac1{\sqrt{2}} (\Phi + \bar\Phi)]_D$ after integration by parts.
In the rigid limit we are considering, the addition of  $-\frac1{\sqrt{2}} (\Phi + \bar\Phi)$ to
the FI operator seems superfluous as it vanishes. However, it is important to keep it, as it shows that the
FI is gauge-invariant independently of the flat supersymmetry background we are working on.
Basically this FI term belongs to the same class as those arising in heterotic string models for anomalous
$U(1)$'s, where the dilaton plays the role played here by $\Phi$ \footnote{This is completely obvious in the deconstructed
version of the model \cite{deconstruct}, where the integrability condition corresponds to the requirement that the FI
be purely along a non-linearly realized $U(1)$, with zero overlap with the linearly realized one.}.
We therefore expect the gauge invariance
of such a FI to be ``robust'' and to resist the extension to supergravity.
Indeed, one can write a bulk operator of the 
form $c(y)[S_0 \bar S_0 (\partial_5 V - \frac1{\sqrt{2}} (\Phi + \bar \Phi))]_D
= c(y)\, e\, (\partial_5 D + \frac i4 \epsilon^{\mu \nu \rho \sigma} 
\bar \psi_\mu \gamma_\nu \psi_\rho F_{\sigma 5} + \cdots )$. 
In this way, the gravitino coupling arising in the supergravity completion of the rigid FI 
term becomes a non-minimal bulk coupling that is perfectly gauge-invariant.

We have seen that FI terms at the orbifold fixed points are quite different from the familiar
FI in 4d. To complete the analysis we want to show that FI terms can be radiatively generated
even in the absence of mixed anomalies, contrarily to what happens in 4d. The calculation of the
one-loop corrections to the FI term is most conveniently done  
working in Euclidean momentum space for the four non-compact directions and in configuration space for the 
fifth dimension, as in \cite{Arkani-Hamed:2001mi}. In this way many subtleties related to the sum
over the KK tower disappear. It is easy to calculate\footnote{No similar tadpole is generated 
for $\Sigma$; this would obviously be non-supersymmetric.}:
\begin{equation}
\label{eq:FI}
\xi(y) = q\,g^2 \int \! \frac {d^4p}{(2\pi)^4} \, (G^+ - G^-)(p;y,y) \;,
\end{equation}
where $G^+$ is the propagator for the even scalar $\phi$ and $G^-$ that for the odd scalar $\phi_c$,
evaluated at points coinciding in the extra dimension.
The propagator for the even field satisfies the equation:
\begin{equation}
\Big[p^2 - \partial_y^2 + m^2 + 2m \big(\delta(y) - \delta(y-\pi R)\big)\Big] G^+(p;y,y^\prime) 
= \frac{1}{2} \left( \delta(y-y^\prime)+\delta(y+y^\prime)\right) \;,
\end{equation}
while the odd propagator satisfies the same equation without boundary mass terms, as it vanishes at the fixed points.
The explicit solutions at coincident points in the fifth dimension are 
for $0\leq y\leq \pi R$ ($\tilde p \equiv \sqrt{p^2 + m^2}$):
\begin{equation}
\begin{split}
G^+(p;y,y) &= \frac 
{\Big[\cosh {\tilde p} y + \frac m{\tilde p}\, \sinh {\tilde p}y \Big]
\Big[\cosh {\tilde p}(\pi R - y) - \frac m{\tilde p}\, \sinh {\tilde p}(\pi R - y)\Big]}
{2\tilde p\,\big(1 - \frac {m^2}{\tilde p^2}\big) \sinh {\tilde p}\pi R} \;, \\
G^-(p;y,y) &= \frac 
{\sinh {\tilde p}y \, \sinh {\tilde p}(\pi R - y)}
{2\tilde p\, \sinh {\tilde p}\pi R} \;,
\end{split}
\end{equation}
with obvious extension to $\pi R\leq y\leq 2\pi R$.

The contributions of the two scalars tend to cancel in the bulk; in fact, we expect a divergent FI 
term to appear only at the boundaries. To study this divergent terms, we can take the limit 
$R \rightarrow \infty$ in $G^+ - G^-$, obtaining:
\begin{equation}
(G^+ - G^-)(p;y,y) \rightarrow
\frac 12 \left[ \frac {e^{-2 \tilde p y}}{\tilde p + m}
+ \frac {e^{-2 \tilde p (\pi R - y)}}{\tilde p - m} \right] \;. 
\end{equation}
We are left with a contribution for each boundary, exponentially suppressed in the bulk.
Since a massive hypermultiplet on $S^1/\Z_2$ has zero modes for any value of its mass, 
it is not possible to regulate the theory \`a la PV; we therefore introduce a momentum 
cut-off $\Lambda$. The FI profile behaves as $\Lambda^3$ at distances 
shorter than $\Lambda^{-1}$. There are quadratic and linear divergences proportional to 
a $\delta$ function and a logarithmic divergence proportional to a $\delta^{\prime\prime}$;
their coefficients can be extracted by using 
$\tilde p\,e^{- 2 \tilde p |z|} \rightarrow \delta(z) + \delta^{\prime\prime}(z)/(4 {\tilde p}^2)$ 
in the UV region $\tilde p \rightarrow \infty$ of the integral. One finds: 
\begin{equation}
\begin{split}
\xi_{\,\text{div}}(y) =
 \frac {g^2\, q}{32 \pi^2}\Bigg\{&\Lambda^2 \Big[\delta(y) + \delta(y - \pi R) \Big]
 - 2\,m \Lambda \Big[\delta(y) - \delta(y - \pi R) \Big] \\
 &+\, \frac 12\, \ln \frac {\Lambda}{|m|} 
\Big[\delta^{\prime\prime}(y) + \delta^{\prime\prime}(y - \pi R) \Big] \Bigg\} \;.
\label{FIm}
\end{split}
\end{equation}
The last term does not correspond to a FI term in the usual sense, but is actually 
a higher-derivative operator.

Consider now the case of several hypermultiplets with charges $q_i$ and masses $m_i$. 
The total $U(1)$-gravitational anomaly is proportional to $\sum_i q_i$. If this anomaly vanishes,
the first term in $\xi(y)$ vanishes too, as it must; otherwise, it would give a FI term in the 4d effective 
theory that is not gauge-invariant in the presence of gravity; the first term is a ``standard'' FI operator.
One is then left with a residual FI term 
satisfying the integrability condition $\int \! dy\,\xi(y) = 0$, which, as we have discussed above, 
corresponds to a gauge-invariant bulk term, whose net effect is to correct the masses of the charged scalars to:
\begin{equation}
\delta m_i = - \frac {g^2}{32\pi^2}\, q_i \sum_j q_j m_j \Lambda \;.
\end{equation}
Note that the radiatively induced FI operators correspond to an odd bulk operator which explicitly
violates the $\Z_2^\prime$ global symmetry discussed in section \ref{odd}. Indeed, this term is
generated only by introducing mass terms that violate this symmetry in the tree-level Lagrangian.

\subsection{The $S^1/(\Z_2 \times \Z_2^\prime)$ model}

Consider now the $S^1/(\Z_2 \times \Z_2^\prime)$ orbifold. In this case no rigid supersymmetry is left
unbroken {\it globally}: each boundary satisfies a different $N=1$ supersymmetry out of the $N=2$. However the
condition that the model be embeddable in 5d supergravity places strong constraints
on the local Lagrangian. For instance, as in the previous section, a hypermultiplet bulk mass term should
be completed by scalar masses localized at the boundaries.
We would like to find the generalization of what we discussed in the 
$S^1/\Z_2$ case, where FI terms can be consistently put on the boundaries without inducing gauge or 
supersymmetry breaking. In this case we have to impose the $D$-flatness condition at each boundary, with 
respect to the residual $N=1$ supersymmetry of that fixed point: in this way locally we have no supersymmetry or gauge 
symmetry breaking; supersymmetry is broken only by non-local terms. The decomposition of the $N=2$ multiplets 
with respect to the two supersymmetries of the two boundaries gives for the auxiliary field of the vector 
multiplet: $D= X_3 -\partial_5 \Sigma$ and $D'= X_3 +\partial_5 \Sigma$, where $X_a$ is the $SU(2)_R$ 
triplet of auxiliary fields of the $D=5$ vector multiplet 
\cite{Mirabelli:1997aj}. The relative minus sign in the expressions for $D$ and $D'$ 
can be understood as a consequence of the $R$-symmetry rotation of the triplet $X_a$, which is needed to go
from an $N=1$ supersymmetry to the other. All the terms $D + \partial_5 \Sigma$ in the Lagrangian (\ref{LKC})
become $D - \partial_5 \Sigma$ with respect to the other supersymmetry, and 
we can thus write explicitly the $D$-flatness conditions at the two boundaries: 
\begin{equation}
\begin{split}
&D = \xi_1\, \delta(y) - \partial_5 \Sigma = 0 \;, \\ 
&D' = \xi_2\; \delta(y- \pi R/ 2) + \partial_5 \Sigma = 0 \;.
\end{split}
\end{equation}
We see that to solve both equations we must have equal (and not opposite as in $S^1/\Z_2$) FI terms on 
the boundaries. As in the $S^1/\Z_2$ case, we expect 
that such FI terms should be easily extended to supergravity: their gauge variation will be
canceled by bulk terms, which explicitly break the local parity symmetry defined in section \ref{odd} for 
the $S^1/(\Z_2 \times \Z_2^\prime)$ case. When the FI terms at the two fixed points are different, (local) supersymmetry 
or gauge symmetry (or both) are spontaneously broken. In this case we expect the supergravity extension to be 
non-trivial and the gravitino charged under the $U(1)$. 

Consider now the one-loop renormalization of the FI term induced by a charged hypermultiplet with scalars
$\phi^{++}$ and $\phi_c^{--}$, whose form is given by an equation analogous to (\ref{eq:FI}). 
As in the $S^1/\Z_2$ case there will be mass terms on the fixed points. At each boundary 
one can use the superfield formalism, as in the previous section, to find the proper term that locally 
restore supersymmetry. 
It is easily found, by using the second term of eq.~(\ref{eq:Vscal}) and the fact that we have to exchange $\phi$ and 
$\phi_c^\dagger$ to go from one rigid supersymmetry to the other, that in this case the two mass terms have the same 
sign at the two fixed points. Therefore we have a scalar potential of the form
\begin{equation}
V_{\rm scalar} = \Big[(m (y) + q \Sigma)^2 (|\phi|^2 + |\phi_{c}|^2) + 2m 
 \sum_{k=0}^3 \delta(y-k\pi R/2)  |\phi|^2 \Big] \;.
\end{equation}
  
The propagator for the odd scalar is again unaffected by the boundary terms, whereas the even propagator satisfies
\begin{equation}
\begin{split}
\Big[p^2 - \partial_y^2 + m^2 + 2m \sum_{k=0}^3 & \delta(y-k\pi R/2) \Big] 
 G^{++}(p;y,y^\prime) = \\ &\frac 14 \sum_{k=0,1} \left( \delta(y-y^\prime+k\pi R)+\delta(y+y^\prime-k\pi R)\right)  \;.
\end{split}
\end{equation}
Their explicit form for coincident points in the extra dimension is, for $0\leq y\leq \pi R/2$
\begin{equation}
\begin{split}
G^{++}(p;y,y) &= \frac 
{\Big[\cosh {\tilde p} y + \frac m{\tilde p}\, \sinh {\tilde p}y \Big]
\Big[\cosh {\tilde p}(\frac {\pi R}2 - y) 
+ \frac m{\tilde p}\, \sinh {\tilde p}(\frac {\pi R}2 - y)\Big]}
{4\tilde p\,\Big[\big(1+\frac {m^2}{{\tilde p}^2}\big) \sinh {\tilde p}\frac {\pi R}2 
+ 2\,\frac {m}{\tilde p}\, \cosh {\tilde p} \frac {\pi R}2 \Big]} \;, \\
G^{--}(p;y,y) &= \frac 
{\sinh {\tilde p}y \,\, \sinh {\tilde p}(\frac {\pi R}2 - y)}
{4\tilde p\, \sinh {\tilde p} \frac {\pi R}2} \;. 
\end{split}
\end{equation}
In the massless case, the result of \cite{Scrucca:2001eb} for the FI profile is easily recovered after 
integrating over the momentum. More generally, one can study the divergent contribution by taking as before 
the limit $R \rightarrow \infty$. The result is obviously the same as in the $S^1/\Z_2$ case because in 
this limit the propagator at one boundary is insensitive to the other boundary. 
The only change is the different sign of the 
mass at the second fixed point: 
\begin{equation}
(G^{++} - G^{--})(p;y,y) \rightarrow \frac 14 \left[
\frac {e^{-2 \tilde p y}}{\tilde p + m}
+ \frac {e^{-2 \tilde p (\frac {\pi R}2 - y)}}{\tilde p + m} \right] \;.
\label{gppgmm}
\end{equation}
The divergent part of the FI terms is then found to be
\begin{equation}
\label{eq:FIlambda}
\xi_{\,\text{div}}(y) =\frac {g^2\, q }{64 \pi^2}\,\left[ \big(\Lambda^2 - 2\,m \Lambda\big) 
\sum_{k=0}^3 \delta(y-k\pi R/2) 
+\, \frac 12\,  \ln \frac {\Lambda}{|m|} \sum_{k=0}^3 \delta^{\prime\prime}(y-k\pi R/2) \right] \;.
\end{equation}
The terms proportional to the $\delta$'s correspond to induced FI terms equal at the two boundaries. 
This, as we said, gives a vacuum that is both (locally) supersymmetric and gauge-invariant. 
The $\delta^{\prime\prime}$ term gives instead a higher-derivative operator.
Note that there is no logarithmic divergence in the $\delta$ terms, although one would naively expect
one on power-counting grounds. The reason is that such a term would correspond to a bulk operator violating the 
local parity, with a coefficient proportional to $m |m|$; but such a term, 
being non-analytic in the mass parameter, cannot be divergent. 

Differently from the $S^1/\Z_2$ case, a theory with a massless Higgs hypermultiplet of charge $q=1$ on 
$S^1/(\Z_2 \times \Z_2^\prime)$ can be regularized with a set of PV fields, as already explained in 
section \ref{pv}. In order to calculate the FI at one loop, the masses and charges of the regulators 
must satisfy $\sum_i q_i=1$ and $\sum_i q_i M_i=0$; this regularizes the 
quadratic and linear divergences
in the $\delta$ terms, and also the logarithmic divergence in the $\delta^{\prime\prime}$ terms.
In the limit $M_i \rightarrow +\infty$, we are left with a divergent contribution of the form
\begin{equation}
\label{eq:FIPV}
\begin{split}
\xi_{\,\text{div}}(y) = 
\frac {g^2}{64 \pi^2} \bigg[ &-2 \sum_i q_i M_i |M_i| \sum_{k=0}^3 \delta(y-k\pi R/2) \\
&+ \frac 14 \sum_i  q_i\ln \frac {|M_i|}\mu \sum_{k=0}^3 \delta^{\prime\prime}(y-k\pi R/2) \bigg] \; ,
\end{split}
\end{equation}
where $\mu$ is an arbitrary renormalization scale. In order to renormalize the theory, we have to add local
counterterms that cancel $\xi_{\rm div}$. After doing so,  
the one-loop correction to the FI term is a {\it UV-finite distribution} $\hat\xi(y)$ which, away from
the boundaries, is purely determined by the 5d massless physical fields. 
Our introduction of a set of PV fields satisfying the two conditions
$\sum_i q_i = 1$, $\sum_i q_i M_i = 0$ and of the local counterterms corresponds to making $\hat\xi(y)$ and $y^2\hat 
\xi(y)$ integrable and finite when $M_i \to \infty$. The integral $\int dy \, y^2 \hat \xi(y) \propto \log \mu R$
is infrared-divergent in the limit of infinite extra dimension: $R$ acts as an IR cut-off.   

Had we chosen a set of PV fields satisfying $\sum_i q_i M_i^2 = 0$ (recall that we must 
take $M_i > 0$ for the regulators to decouple), the FI terms would have been finite.  
We will see in section \ref{SUGRA} that 
this particular choice of regularization has a clear interpretation in the 5d supergravity extension 
of the theory.  
It is worth noting that, with this choice, the zero mode of $\hat\xi$ (i.e. its integral over the 
full circle) vanishes exactly, while we would naively have expected a non-zero answer scaling like $1/R^2$.
This result is less surprising when the calculation is done through a KK mode expansion.
In that case, the zero mode of $\xi$, before regulation, is determined by just one diagram
where the scalar zero mode circulates; it is therefore formally independent of $R$, although infinite.
The regulated result then should be finite while being still $R$-independent, so it must
be zero.

\section{\label{SUGRA}Supergravity embedding}

In models where the existence of a locally conserved supercurrent is crucial (as in \cite{Barbieri:2000vh}) 
to constrain the possible 
divergences appearing in the effective action, one is naturally led to consider a supergravity embedding.
It is therefore important to show how what we said in the previous sections
is explicitly extended to 5d supergravity, in particular the 
introduction of a gauge-invariant PV regulator with a stepwise mass term.

Every locally supersymmetric 5d theory must contain the gravity multiplet, composed by the graviton $g_{MN}$, 
a symplectic Majorana gravitino $\psi_M$ and a vector field $A_M$: the graviphoton. If we add to the pure
gravitational theory $n_V$ vector multiplets, the Lagrangian is completely 
determined by the geometry of the manifold ${\cal M}$ parametrized by the $n_V$ vector-multiplet scalars $\phi^x$. 
This geometry is specified starting from a $(n_V+1)$-dimensional 
space ${\cal C}$ with coordinates $b^i$ and K\"ahler potential ${\cal K}(b^i) = d_{ijk} b^i b^j b^k$. 
The scalar manifold ${\cal M} \subset {\cal C}$ is then defined through the constraint ${\cal K}(b^i)=6$ 
on which $b^i = b^i(\phi^x)$ and the metric $G_{ij} = -\frac 12 \partial_i \partial_j \ln {\cal K}(b^i)$ 
on ${\cal C}$ induces a metric $G_{xy} = \partial_x b^i \partial_y b^j G_{ij}$ on ${\cal M}$
\footnote{The origin of this nice geometric description lies in the fact that 5d supergravities can 
be obtained as compactifications of eleven-dimensional supergravity on a Calabi--Yau space. 
The symmetric tensor $d_{ijk}$ appearing in the K\"ahler potential encodes the intersection 
numbers of the Calabi--Yau and the constraint ${\cal K} = 6$ corresponds to constant volume, whose
fluctuations are described by an independent universal scalar field. See for example \cite{Lukas:1998tt}.}. 
The Lagrangian is \cite{Gunaydin:1983bi}:
\begin{equation}
\begin{split}
\label{eq:SUGRAL}
{\cal L} = 
-\frac e{2\,\kappa^2} \Big[&R + G_{xy}\, \partial_M \phi^x \partial^M \! \phi^y + G_{ij}\, F^i_{MN} F^{j\,MN} 
+ \frac {d_{ijk}}{6\sqrt{2}}\, \epsilon^{MNPQR}\, A^i_M F^j_{NP} F^k_{QR}  \\
& - \frac {i}{2\sqrt{2}} \big(\bar \psi_M \Gamma^{MNPQ} \psi_N 
+ 2\,\bar \psi^P \psi^Q \big) \, b_i F^{i}_{PQ} 
+ \cdots \Big] \;,
\end{split}
\end{equation}
where the dots stand for terms involving the gauginos, which will not play any role in our discussion,
and four-fermion terms. We indicate with $A^i$ the set of $n_V$ vectors and the graviphoton. 
Note that Chern--Simons terms are a fundamental ingredient 
of 5d supergravity. 

Matter hypermultiplets can be easily introduced in the theory. Interactions with the vector multiplets 
can be achieved by gauging suitable isometries of the hypermultiplet scalar manifold
(for a complete discussion see \cite{Ceresole:2000jd} and references therein
\footnote{For a $D=4$, $N=2$ pre-geometric approach, see \cite{Zachos:iw}.}). 
A similar gauging is necessary also to generate 
mass terms and a potential for the hypermultiplets. In particular the hypermultiplet mass terms of rigid 
supersymmetry follow here from a gauging with the graviphoton: the mass of each hypermultiplet is 
proportional to its gravicharge. We are interested in a flat gauging, in which no cosmological 
constant is generated. It is well-known that this is possible, provided that the gravitino has zero charge
or, in other words, that the gauged isometry is not part of the $R$-symmetry. 

Consider now such a supergravity theory on an orbifold. Under any orbifold symmetry, the bosonic fields 
$(g_{\mu\nu}, g_{55}, A_5)$ of the gravity multiplet must be taken even, while $(g_{\mu 5}, A_\mu)$ are 
odd. If we admit only even operators, the theory remains locally supersymmetric, both in the bulk and at 
the fixed points. Odd operators can be consistently introduced along the lines of \cite{Bergshoeff:2000zn}.
To obtain a stepwise coefficient $g$, one promotes it  
to a field $G(x)$, invariant under supersymmetry: in this way the Lagrangian is no longer supersymmetric 
and its variation may be written as $\delta {\cal L} = J_M  \partial^M G$. 
This variation can be canceled by introducing a new four-form field $A_4$ with Lagrangian 
${\cal L}_{\rm mult} = A_4 \wedge d G$ and transforming as $\delta A_4 = - {}^*J$, in such a way that 
$\delta ({\cal L} + {\cal L}_{\rm mult}) = 0$. The field $A_4$ acts as a Lagrange multiplier forcing 
$\partial_M G = 0$ through its equation of motion, which yields back a constant $G$ on-shell, 
i.e. the situation we started with. 
At this stage, however, a non-trivial generalization can be obtained by introducing sources for $A_4$. 
Adding terms of the form ${\cal L}_{\rm source} = \sum_i \mu_i \delta(y - y_i) (A_4 + W)$, 
where $W$ is an additional term that is required to make $\delta {\cal L}_{\rm source} = 0$ on its own,
the equation of motion of $A_4$ yields $\partial_5 G = \sum_i \mu_i \delta(y - y_i)$, which implies that 
$G$ must have a piecewise constant profile on-shell. 
Notice that this extension is consistent provided the integrability condition $\sum_i \mu_i = 0$ is satisfied. We
have thus described a general method to introduce odd operators in supergravity, which, after the 
elimination of the Lagrange multiplier 4-form, reduces to a proper addition of operators at the fixed 
points, which makes the full action invariant. 

At this point, we are ready to discuss the generalization of the results obtained in the previous sections.
From the above reasoning, it is clear that there is no obstruction to the introduction of a PV regulator with a
jumping mass. The boundary mass terms found in section \ref{FI} arise here from the 
$W$ term, which is needed to make the fixed-point action invariant. This means that, even in supergravity, it is
possible to regularize the theory in a gauge-invariant way, so that no anomaly will arise. As in the rigid
limit, the PV regulator introduces a local CS term in the limit of infinite mass, with a
jumping coefficient that is again described along the lines of \cite{Bergshoeff:2000zn}. 
Since the PV field has a non-vanishing gravicharge $r_i$ related to its mass, $M_i \propto r_i \kappa^{-1}$, 
a CS term with a photon and two graviphotons will also be generated, with a coefficient proportional to 
$\sum_i q_i r_i^2$.

In rigid supersymmetry we have seen that we can add FI terms on the boundaries with an arbitrary 
coefficient, modifying in this way the VEV of the scalar $\Sigma$. However, we also noticed that the FI
should truly be regarded as an odd bulk operator, since in this way its full gauge invariance is
manifest. We have also seen that in supergravity odd bulk operators are made consistent
only by the addition of specific boundary terms \cite{Bergshoeff:2000zn}. It is then natural
to expect that in supergravity the FI term is completely fixed by the bulk Lagrangian, i.e. by
the structure constants $d_{ijk}$. Note that since the FI fixes the VEV profile of the gauge scalars $\Sigma$,
then the structure constants $d_{ijk}$ themselves will determine where on the manifold ${\cal M}$ the vacuum 
is.

We will now explain this in more detail. In practice we will check that the additive renormalization
of the hypermultiplet mass term, which is generated in the flat theory by the FI term, can also be 
calculated in the supergravity extension just by considering the $d_{ijk}$.
Let us focus on  a supergravity model with $n$ hypermultiplets and with a single vector field 
$B_M$ (which plays the role of hypercharge
in the model of ref. \cite{Barbieri:2000vh}) 
besides the graviphoton $A_M$. We consider the interesting situation where
$B_\mu$ is even under the orbifold symmetry, while $A_\mu$ is necessarily odd.
The two vectors gauge
a $U(1)_A\times U(1)_B$ isometry of the hypermultiplet manifold, which we assume to be linearly realized.
Around the $U(1)^2$ symmetric point, the action of the two isometries is just described in the linearized
approximation by two commuting charge matrices $Q_A(y)=\eta(y)\hat Q_A$ and $Q_B$ acting on the $n$-dimensional 
hypermultiplet.
Notice that since $A_\mu$ is odd its charge matrix undergoes an overall sign change at each fixed point 
\cite{Bergshoeff:2000zn}. Five-dimensional supergravity fixes the hypermultiplet mass matrix to be 
$M(Q_Ab^A(\phi)+Q_Bb^B(\phi))$ where $M$ is just the 5d Planck scale up to some fixed constant (which is however
unimportant in the following discussion). Notice that since the mass terms are odd operators, $b^A$ must
be even and $b^B$ must be odd. Without loss of generality we
 can then choose $b^B=\phi/M$ where $\phi$ is an odd scalar, while $b^A$ is fixed in terms of $\phi$
through the constraint $d_{ijk}b^ib^jb^k=6$. The scalar $\phi$ is the superpartner of $B$ when we take the flat 
limit. Now, let us start from
a Lagrangian without parity-odd operators in the pure gauge sector: only the constants $d_{AAA}=6$ 
and $d_{ABB}=1/(Mg_B^2)$ will be different from zero. Here we put in evidence the relation with the
$U(1)_B$ coupling $g_B^2$.
Since there are no odd terms in the gauge sector, following the procedure
of ref.~\cite{Bergshoeff:2000zn} we should not add any boundary term to enforce local supersymmetry.
In particular, there should be no boundary term involving just $\phi$, and since $\phi$ is odd
it must vanish on the vacuum. Then the solution of the structural constraint will just
be $b^A=1$, $b^B=0$, and the matter mass matrix will point along the graviphoton charge.
What we have just outlined is the supergravity picture of a rigid theory without $U(1)_B$ FI term.
The absence of a FI term is also consistent with the fact that the pure gauge supergravity Lagrangian
is quadratic in $B$: $d_{AAB}=0$. Consider now a slight deformation
of the previous model where we introduce a (small) $d_{AAB}=2\epsilon \eta(y)d_{ABB}$.
The new term introduces a kinetic mixing between $B$ and $A$ along with an $AAB$ CS term.
Now we have terms that jump at the boundary and we must be careful to work out the boundary Lagrangian according to
\cite{Bergshoeff:2000zn}. However, one can straightforwardly map the deformed theory back to the
original one by noticing that through the field redefinition $A_M=\tilde A_M$, 
$B_M=\tilde B_M-\epsilon \eta(y)\tilde A_M$, the structure constants go back (up to $O(\epsilon^2)$
terms) to the original ones, with $d_{AAB}=d_{BBB}=0$. (Notice that the field strengths rotate precisely as $A$
and $B$, without spurious terms,  in spite of the presence of $\eta(y)$. This is thanks to $A_M$ being odd;
a field redefinition of the type $A_M\to A_M+\eta(y) B_M$ would not be well behaved at the boundaries.)
In the tilded basis we must have the same result as before: $\tilde b^A=1$, $\tilde b^B=0$, and by transforming
back we find $b^A\simeq 1$, $b^B=-\epsilon \eta(y)$. The mass term of the deformed theory is 
therefore $M(Q_A -\epsilon \eta(y)Q_B)$: a new term along $Q_B$ has been generated. The case we have just described
should correspond to the supergravity extension of a flat theory with a FI $\propto d_{AAB}$. We have 
explicitly checked this by considering a theory where $d_{AAB}=0$ at tree level and by calculating the 
one-loop renormalizations of $A\wedge d A\wedge d B$ and of the $B$ FI, which are both $\propto{\rm Tr}(Q_AQ_AQ_B)$.
We find that the correction to the matter mass matrix calculated with the above supergravity reasoning
coincides with the direct calculation in rigid supersymmetry via the FI term.
This explicitly shows that the FI term is just the flat limit remainder of a non-zero $d_{AAB}$.  

An independent argument for this interpretation can be obtained by looking at the gravitino term 
$\epsilon^{\mu \nu \rho \sigma} \bar \psi_\mu \gamma_\nu \psi_\rho F_{\sigma 5}$ appearing in the $N=1$ 
supergravity completion of gauge-invariant FI terms (see section \ref{z2}), which in the $N=2$ theory must come
from the last term in the Lagrangian (\ref{eq:SUGRAL}). As we have seen, a vanishing FI term corresponds to only 
$b_A$ different from zero so that the gravitino couples only to the graviphoton. A non-vanishing FI term leads 
instead to $b_B \neq 0$, implying a coupling between the gravitino and the corresponding field strength.

Summarizing, FI terms are seen as part of a parity-odd $N=2$ supergravity bulk operator, which contains a
CS term with the corresponding gauge field and two graviphotons. If we cannot regulate the theory 
maintaining the local parity symmetry, these terms will be radiatively generated. The advantage of the supergravity
point of view is that we can better understand the condition $\sum_i q_i M_i^2 = 0$: the absence of FI terms is equivalent
to a choice of regularization such that a CS term for two graviphotons and a photon is not generated or,
equivalently, which does not mix the two vectors. We stress, however, that from the effective field theory 
viewpoint, there is no symmetry we can impose to avoid such operators, the only possible candidate,  5d parity,
being broken to ensure gauge invariance.

\section{\label{conclusions} Conclusions}

We have shown that orbifold field theories with a non-anomalous spectrum of zero modes can
in general be made gauge-invariant only at the price of introducing odd operators. In particular there should effectively 
be a Chern--Simons
term with a coefficient unambiguously fixed by the requirement of anomaly cancellation.
 These operators are allowed by the orbifold symmetry,
but violate  local 5d parity. The introduction of these new interactions has in general an important
phenomenological impact, chiefly for non-supersymmetric models. Indeed 5d parity plays the role of a discrete
chiral symmetry; by its explicit breaking Dirac fermion masses can no longer be naturally forbidden.
For instance, at the two-loop level, the CS term gives rise to a quartically divergent mass for charged fermions.
In supersymmetric models such as \cite{Barbieri:2000vh}
bulk supersymmetry limits the appearance of odd operators; mass terms for the hypermultiplets are however allowed:
in particular a tree-level mass term could be added for the Higgs multiplet.  
If tree-level masses are set to zero, the non-renormalization theorem guarantees that they can only
be renormalized through the generation of FI terms \cite{Ghilencea:2001bw,Barbieri:2001cz} 
and by finite, small,  non-local corrections associated
to the orbifold breaking of supersymmetry. Indeed the on-shell effect of a FI term in the limit
of rigid supersymmetry is the same as a shift in the VEV of the scalar fields in the vector $N=2$ multiplets:
it shifts the hypermultiplet masses. In the supergravity embedding a FI for a vector $B_M$ is associated to the presence
of a CS term of the type $A\wedge dA\wedge dB$, where $A_M$ is the graviphoton.  This fact
 suggests that the FI  renormalization vanishes beyond one-loop. This is because CS terms
are in turn associated to anomalies and cannot be renormalized beyond one-loop. Moreover,
the one-loop renormalization of the FI term is given by the $B$ and graviphoton charges through
a suggestively simple algebraic relation: $\xi\propto {\rm Tr}(Q_AQ_AQ_B)\propto d_{AAB}$. 
So while from pure symmetry considerations it is not possible to argue in favour of a vanishing FI term,
nonetheless its form and renormalization properties are so restricted that it does not seem unreasonable
to single out the case $\xi =0$ for phenomenological study. 
Of course a satisfactory explanation of $\xi=0$
can only come from a more fundamental theory. One possible direction is to try to obtain our 5d model as
the low energy limit of 11d supergravity compactified on a Calabi--Yau, then the $d_{ijk}$ and thus the FI
are calculable topological quantities.

\section*{Note Added}
During the completion  
of the paper, ref.~\cite{pirati} appeared, which has some overlap with our section \ref{pv}.
However, we disagree with the authors on their conclusion that the addition 
of odd operators spontaneously breaks the $\Z_2 \times \Z_2^\prime$ orbifold symmetry.

\section*{Acknowledgments} 
We are indebted with V.I.~Zakharov for his helpful collaboration in the early stage of this work.
We also thank G.~Dall'Agata, J.P.~Derendinger, S.~Ferrara, A.~Hebecker, M.~Serone, L.~Silvestrini
and F.~Zwirner for useful discussions. This work has been partially supported by the EC under
TMR contract HPRN-CT-2000-00148.

%%%%%%%%%%
%%%%%%%%%%      Appendix
%%%%%%%%%%

\appendix

\section*{Appendix}

\section{\label{regulators}Two different regulators}

We give here the details of the computation of the anomaly exhibiting two
different
regulators for the three-point function. The first one is a higher
derivative
deformation of the fermionic kinetic term, which breaks gauge invariance,
preserves
the local parity defined in section~\ref{pv} and reproduces the result of
eq.~(\ref{anomaly}); the second one is the massive gauge invariant
Pauli--Villars, which leads to a non-anomalous theory with local parity
explicitly broken.
The Lagrangian for a 5d fermion on $S^1/(\Z_2\times \Z^\prime_2)$ with an arbitrary 
jumping mass term $M\, \eta(y)$ is
\begin{equation}
\label{Ltot}
{\cal L} = \int_0^{2\pi R}\!\!\! dy\, \left[-\frac{1}{4g^2} F^{MN}F_{MN}+\bar \psi\,
i\Gamma^{M} \big(\partial_M -iA_M\big) \psi- M \eta(y) \bar\psi \psi\right] \;.
\end{equation}
Using the decomposition of eqs.~(\ref{decompPsi}) and (\ref{decompA}), this gives:
\begin{equation}
\begin{split}
{\cal L} =& \sum_{k} \bar \psi_k\
 (i\!\!\not\! \partial -M_k)\ \psi_k
 +\sum_{k,l,n} \alpha^{+++}_{k\, l\, 2n}
 \left[ P_n^+\ \bar \psi_k \gamma^\mu A_\mu^{2n} \psi_l +
        P_n^-\ \bar \psi_k \gamma^\mu \gamma^5 A_\mu^{2n} \psi_l \right]
\\
 &+\sum_{k,l,n} \alpha^{+--}_{k\, l\, 2n}
 \left[ P_n^+\ \bar \psi_k\ i A_5^{2n} \psi_l -
        P_n^-\ \bar \psi_k\ i \gamma^5 A_5^{2n} \psi_l \right] \;,
\end{split}
\end{equation}
where $P_n^{\pm} = \frac 12 (1 \pm(-1)^n)$ and the overlaps $\alpha^{\pm\pm\pm}_{k\,l\, 2n}$
are defined by
\begin{equation}
\alpha^{\pm\pm\pm}_{k\, l\, 2n} = \int^{2\pi R}_0 \!\!\! dy\
\xi^{\pm\mp}_k(y)\,\xi^{\pm\mp}_l(y)\, \zeta^{\pm}_{2n}(y) \;.
\end{equation}
The divergence of the currents is found to be
\begin{equation}
\label{corrente}
\begin{aligned}
\left(\partial_\mu J^\mu\right)^{2n} &=
 \sum_{k,l} \alpha^{+++}_{k\, l\, 2n}
 \left[ P_n^+ \partial_\mu\left(\bar \psi_k \gamma^\mu \psi_l \right) +
        P_n^- \partial_\mu\left(\bar \psi_k \gamma^\mu \gamma^5
\psi_l\right) \right] \\
\left(\partial_5 J^5 \right)^{2n} &=
 \sum_{k,l} i \left(M_l+(-1)^{n+1} M_k \right) \alpha^{+++}_{k\, l\, 2n}
 \left[ P_n^+ \bar\psi_k  \psi_l - P_n^- \bar\psi_k \gamma^5 \psi_l
\right] \;.
\end{aligned}
\end{equation}
The wave functions satisfy eq.~(\ref{prop}) and are given by
\begin{equation}
\label{funzonda}
\begin{split}
\xi^{+-}_k(y) &= \frac{1}{\sqrt{\pi R}} \left(1-\frac{\sin M_k \pi R}{M_k \pi R}\right)^{-1/2} \times
\begin{cases}
\sin M_k (y-\pi R/2)  & y\in [0,\pi R] \\ 
\sin M_k (3\pi R/2-y) & y\in [\pi R,2\pi R] 
\end{cases}  \\
\xi^{-+}_k(y) &= \xi^{+-}_k(y+\pi R/2) \;,
\end{split}
\end{equation}
the mass eigenvalues $M_k$ being defined by eqs.~(\ref{implicit}) and (\ref{eigenvalue}).
For each $k$ there is a 4d fermion $\psi_k$ with mass
$M_k=\sqrt{\lambda_k^2+M^2}$.
In the limit $M\to 0$ we recover the massless wave functions of
eq.~(\ref{masslesswf})
and the usual
spectrum, while for $M\to +\infty$ the whole tower of states becomes
infinitely massive
and decouples. When $M<-2/(\pi R)$ one right-handed and one left-handed
state appear localized respectively on $y=0,\pi R/2$ fixed points,
becoming
massless in the limit $M\to -\infty$.
This means that only a fermion with positive mass can play the role of a
PV regulator
\footnote{What is actually relevant is the relative sign between the mass
and $\Gamma^5$.
Indeed, in five dimensions there are two inequivalent
representations for $\Gamma^5$ which differ by a sign. Changing this
convention
also changes the sign of $M$ which gives localized states, but of course
the
physics remains the same.}.

Now that we have shown the basic formulae, we introduce the first
regulator, which
consists in replacing the fermion kinetic term in eq.~(\ref{Ltot}) with
\begin{equation}
\label{ZJreg}
\big( i\!\not\! \partial - M\eta(y)\big) \to \big( i\!\not\! \partial - M\eta(y)\big)\,
e^{\epsilon \big(i\,\not \partial + M\eta(y)\big)\big( i\,\not \partial -
M\eta(y)\big)} \;,
\end{equation}
where $\not\!\! \partial = \partial_M\, \Gamma^M$. The wave functions
$\xi^{\pm\mp}$ still
satisfy eq.~(\ref{prop}) and the propagator of each KK mode $\psi_k$
gets an exponential
convergence factor $\exp\epsilon(q^2-M_k^2)$ in momentum space.
It is evident that this choice explicitly breaks 5d gauge invariance, even though
it is even under the local parity defined in section~\ref{pv}.
This procedure regulates both the 4d loop integral and the series over the
KK modes
in the three-point function for a 5d fermion with (arbitrary) jumping
mass.
More in detail, working with the KK mode expansion, different diagrams 
can be formed when we consider the insertion of $\partial_\mu J^\mu$,
$\partial_5 J^5$
in the background of two gauge bosons.
Depending on the external 5d momenta $n_1,n_2,n_3$ (see fig.~\ref{diag}), any of the
three vertices can be either vectorial, or axial, or scalar, or pseudoscalar.
However, as previously stated, the sum of the two diagrams of fig.~\ref{diag}
with an even number of $\gamma^5$ is odd under the charge conjugation of
eq.~(\ref{chargeC})
and therefore vanishes.
For the remaining diagrams, the external momenta are such that
$n_1+n_2+n_3$ is odd, corresponding to an operator in the effective action
which violates the conservation of momentum in the fifth direction and as
such can only be a boundary operator.
Dimensional analysis tells us that the naive divergence of these diagrams
will be at most linear (and not quadratic), as expected for a boundary
operator or, in other
words, the sum over KK modes is convergent. This is clear in the massless
limit where, if $n_1+n_2+n_3$ is odd, only a finite number of KK modes circulate in the loop.
It is clear from these considerations that there will be no contribution
 from the background $A_5$, as it vanishes on the boundary.
The only remaining diagrams are of the type $AVV$, $AAA$, $VAV$ for
$\langle\partial_\mu J^\mu\rangle$ and $PVV$, $PAA$, $SAV$ for
$\langle\partial_5 J^5 \rangle$ (where the first label refers to the
current).

Let us consider the graph $AVV$ with external 4d momenta $k_{1\,\alpha}$,
$k_{2\,\beta}$, fixed modes $k,l,m$ running in the loop and 4d loop momentum 
$p$, as depicted in fig.~\ref{diag}. Using the identity 
$(\not\! k_1\,+\!\not\! k_2)\gamma^5= (\not\!p\;+\not\!k_1+M_k)\gamma^5-
(\not\! p\;-\not\! k_2-M_l)\gamma^5-(M_k+M_l)\gamma^5$,
the trace over Dirac matrices can be split into three terms
\begin{equation}
\label{trace}
\begin{split}
\Tr \bigg[(\not\! k_1\, &+ \not\! k_2) \gamma^5 \frac{1}{\not\! p\;+\not\!
k_1-M_k} \gamma^\alpha
\frac{1}{\not\! p\,-M_m} \gamma^\beta \frac{1}{\not\! p\;-\not\! k_2-M_l}
\bigg] +
 \text{symm.} = \\
&-(M_k+M_l) \bigg\{
\Tr \bigg[ \gamma^5 \frac{1}{\not\! p\;+\not\! k_1-M_k} \gamma^\alpha
 \frac{1}{\not\! p\,-M_m} \gamma^\beta \frac{1}{\not\! p\;-\not\!
k_2-M_l}\bigg]
 +\text{symm.} \bigg\} \\
&-\bigg\{\Tr\left[\gamma^5\gamma^\alpha\frac{1}{\not\! p\,-M_m}\gamma^\beta
\frac{1}{\not\! p\;-\not\! k_2-M_l}\right]+\text{symm.}\bigg\} \\
&-\bigg\{\Tr\left[\gamma^5\frac{1}{\not\! p\;+\not\! k_1-M_k}\gamma^\alpha
\frac{1}{\not\! p\,-M_m}\gamma^\beta\right]+\text{symm.}\bigg\} \; .
\end{split}
\end{equation}
The first term converges when integrated over $p$, even if one removes the
regulator, and
gives a contribution exactly equal and opposite to the $PVV$ insertion of
$\partial_5 J^5$.
The other two, instead, would diverge linearly without the regulator, even
if
their sum is convergent.  The presence of the regulator is then crucial to
define
each term separately by removing the routing ambiguity. More explicitly,
using the regularization recipe of eq.~(\ref{ZJreg}), the sum of the last
two terms in
eq.~(\ref{trace}) gives an
anomalous piece $\sim \epsilon^{\alpha\mu\beta\nu} k_{1\,\mu} k_{2\,\nu}$
which does not
depend on the mass of the modes, in the same way as happens in 4d.
Similar considerations apply for the other diagrams $AAA$, $VAV$ and
$PAA$, $SAV$
and resumming over all KK modes one finds the result of
eq.~(\ref{anomaly}).
We stress that the anomaly we found does not depend on the value of the mass
$M$
\footnote{This extends the result of ref.~\cite{Scrucca:2001eb} and shows
that what
found in \cite{Arkani-Hamed:2001is} holds true also for the
$S_1/\Z_2\times\Z_2^\prime$ orbifold.}
and this is an indication that a massive fermion could act as a regulator
itself.

Let us now discuss the Pauli--Villars regularization.
In order to be a regulator for the theory with a massless fermion, the
massive PV must resolve
the ambiguity coming with the last two terms in eq.~(\ref{trace}).
This is exactly what happens. An arbitrary shift of the dummy variable in
a linearly divergent
integral corresponds to a surface finite contribution;
in our case, a shift of the loop momentum $p\to p+a$ in the integral of
the last two terms in
eq.~(\ref{trace}) gives a finite term that does not depend on the KK
masses:
\begin{equation}
\Delta^{\alpha\beta}(a) = 2i\pi^2 \frac{a^\mu}{(2\pi)^4} \lim_{p\to\infty}
p^2 p_\mu
 \frac{\Tr\left(\gamma^5\gamma^\alpha\not\! p\gamma^\beta (\not\!
k_1+\not\! k_2)\right)}{p^4}
 = \frac{1}{8\pi^2} \epsilon^{\alpha\mu\beta\nu} a_{\mu} (k_1+k_2)_\nu \; .
\end{equation}
Summing over all KK modes, one gets an arbitrary piece in the $AVV$
amplitude
proportional to the sum of the overlap integrals
\begin{equation}
I_{n_1 n_2 n_3} = \sum_{k,l,m} \alpha^{+++}_{k\,l\,2n_1}
\alpha^{+++}_{k\,m\,2n_2} \alpha^{+++}_{l\,m\,2n_3} \; .
\end{equation}
Using the completeness relations for our basis of wave functions, without
ever using their expression, it is not difficult to show that $I_{n_1 n_2 n_3}$ does
not depend on the mass $M$ but only on the external modes; more precisely
\footnote{When the external modes satisfy momentum conservation,
$n_1+n_2=n_3$, then $I_{n_1 n_2 n_3}$ diverges and indeed in this case
an infinite number of KK modes circulate in the loop.}
\begin{equation}
I_{n_1 n_2 n_3} = \frac{a_{2n_1}a_{2n_2}a_{2n_3}}{2} \quad\quad\quad
 \text{for} \quad n_1+n_2+n_3 \;\text{odd.} 
\end{equation}
From this relation, it follows that the shift in the amplitude cancels when
the PV contribution is added to the physical (massless) one or,
in other words, that the leading divergence of the sum of the PV modes
matches exactly the leading divergence of the physical diagram. We conclude that
a massive 5d fermion with a positive jumping mass regulates the massless
theory, removing any ambiguity.

We have shown that the regulator (\ref{ZJreg}) breaks the 5d gauge
invariance, and
this causes an anomalous term to appear. The 5d PV, on the contrary, preserves the
gauge symmetry and the sum of the last two terms in eq.~(\ref{trace}), regulated by the
corresponding PV contribution, now vanishes. 
Again, the first term in eq.~(\ref{trace}) is canceled by the $PVV$ insertion of $\partial_5 J^5$,
analogously to what we found with the regulator (\ref{ZJreg}).
The same happens with $AAA$, $PAA$, and $VAV$, $SAV$, so that $\partial_M J^M=0$ and the theory is non-anomalous.

\section{\label{fuji} Fujikawa's approach and index theorems}

In this appendix we will show how one can compute anomalies in 
a 5d orbifold theory using Fujikawa's method \cite{Fujikawa:1979ay}, in
which anomalies are interpreted as the variation of the measure for fermions 
in the functional integral for the effective action, making the arguments outlined in \cite{Scrucca:2001eb}
more precise.

We consider the $U(1)$ gauge transformations of a 5d Dirac spinor $\psi$:
\be
\psi \rightarrow e^{i \alpha(x)} \psi \;,\;\;
\bar \psi \rightarrow e^{-i \alpha(x)} \bar \psi \;.
\end{equation}
The variation of the measure under an infinitesimal transformation
of this type is:
\be
{\cal D}\psi {\cal D}\bar \psi \rightarrow  
\exp \Big[-i\int \! d^5x \, \alpha(x)\, 
({\rm Tr}_{\{\psi_n\}}\, 1\hspace{-4pt} 1 
- {\rm Tr}_{\{\bar \psi_n\}}\, 1\hspace{-4pt} 1)\Big] 
{\cal D}\psi {\cal D}\bar \psi \;,
\label{jacob}
\end{equation}
where $\{\psi_n(x)\}$ and $\{\bar \psi_n(x)\}$ are the eigenfunctions of $D^\dagger D$
and $D D^\dagger$, where $D$ is the Dirac operator. On the circle, $D = D^\dagger$
\footnote{We are implicitly assuming an analytic continuation to 
Euclidean space, to have a Hermitian Dirac operator.}, contrarily to what happens 
for a 4d Weyl spinor and therefore the anomaly trivially vanishes.

Consider now the case of an orbifold projection. The traces in 
(\ref{jacob}) must then be restricted to those states left invariant 
by the projection, by inserting the proper projection operator. 
Whenever the projection acts differently on $\psi$ and $\bar \psi$, the 
anomaly will no longer vanish. In the simplest case of $S^1/\Z_2$ generated
by a reflection $R : y \rightarrow -y$ acting chirally on the fermions,
\be
\psi(-y) = \gamma^5 \psi(y) \;,\;\; 
\bar \psi(-y) = - \bar \psi(y) \gamma^5 \;,
\end{equation}
eq.~(\ref{jacob}) becomes:
\be
{\cal D}\psi {\cal D}\bar \psi \rightarrow  
\exp \Big[-i\int \! d^5x \, \alpha(x)\, 
\Big({\rm Tr}_{\{\psi_n\}}\, \frac {1 + Q}2 
- {\rm Tr}_{\{\bar \psi_n\}}\, \frac {1 - Q}2\Big) \Big] 
{\cal D}\psi {\cal D}\bar \psi \;,
\label{eq:jacob2}
\end{equation}
where $Q=R \gamma^5$. Since the two sets $\{\psi_n\}$ and $\{\bar \psi_n\}$ are 
equivalent, one finally finds an anomaly given by
\begin{equation}
\label{eq:trace}
{\cal L}(x) \to {\cal L}(x)+ \alpha(x) A(x) \, ;\qquad A(x) = - {\rm Tr}_{\{\psi_n\}}\,Q \;.
\end{equation}
As in the standard case, it is possible to give an index interpretation 
of this expression. The key ingredient to do so is that $Q$ satisfies 
$\{D,Q\}=0$ and $Q^2=1$. Since $[D^2,Q] = 0$, the two operators can be 
simultaneously diagonalized and one can show that only zero modes of $D^2$ 
contribute to ${\rm Tr}\,Q$. In fact, for any non-zero mode $\psi_n$
of $D^2$, there exists another mode, $D \psi_n$, with the same $D^2$ 
eigenvalue but opposite $Q$. Therefore we can rewrite the integral of 
eq.~(\ref{eq:trace}) as
\be
\label{eq:zeromode}
\int d^5x A(x) = - \left( n_{Q=+}^{D^2=0} - n_{Q=-}^{D^2=0} \right) \;.
\end{equation}
We can write this expression as an explicit index because if we go to the basis that 
diagonalizes $Q$ we have:
\be
Q = \begin{pmatrix} 1 \hspace{-4pt} 1 & 0 \cr 0 & - 1 \hspace{-4pt} 1 \end{pmatrix}  
\;,\;\; D = \begin{pmatrix} 0 & D_+^\dagger \cr D_+ & 0 \end{pmatrix} \;,\;\; 
D^2 = \begin{pmatrix} D_+^\dagger D_+ & 0 \cr 0 & D_+ D_+^\dagger \end{pmatrix} \;, 
\end{equation}
so that eq.~(\ref{eq:zeromode}) can be rewritten as a function of ${\rm Ker}\, D_+$:
\be
\int d^5x A(x) = - ({\rm dim}\,{\rm Ker}\, D_+ - {\rm dim}\,{\rm Ker}\, D_+^\dagger) 
= - {\rm index} \; D_+ \;.
\end{equation}

We can go further and express the integrated anomaly in terms of the gauge connection, as in the usual
4d case. To do so it is useful to relate the trace of eq.~(\ref{eq:trace}) to the trace over $\gamma^5$ restricted
to the fixed points of $R$:
\be
\label{eq:5D4D}
{\rm Tr}_{\{\psi_n\}}\,Q = \frac 12 \left[\delta(y) + \delta(y - \pi R)\right]\, 
{\rm Tr}_{\{\psi_n\}} \gamma^5 \;.
\end{equation}
To obtain this relation, we regularize the trace in an explicitly gauge-invariant way through
the 5d Dirac operator:
\begin{equation}
\begin{split}
\label{eq:trreg}
A(x) =\;& - \sum_k \psi_k^\dagger (x,y) Q \psi_k(x, y) = - \sum_k \psi_k^\dagger (x,y) \gamma^5 \psi_k(x, -y)\\ 
=& - \lim_{M \rightarrow \infty} \lim_{y \rightarrow x}
{\rm Tr} \, \gamma^5 \, {\rm exp}[-(D/M)^2] \, \delta(x-y) \, [\delta(2y) + \delta(2 y - 2\pi R)] \,.
\end{split}
\end{equation}
Note that with respect to the standard 4d case we have an additional piece in the Dirac operator:
$\Gamma^5 (\partial_5 - i A_5)$. Actually this extra piece does not contribute because the trace is non-zero
only when we take $\gamma^5$ with four 4d gamma-matrices coming from the expansion of the exponential. We thus obtain
\be
\label{eq:fin}
A(x) = \frac12 \left[\delta(y) + \delta(y-\pi R)\right] \frac {-1}{32\pi^2} F_{\mu\nu}\tilde F^{\mu\nu} \;. 
\end{equation} 
Note that this result differs by a factor of 3 from what we obtained in the evaluation of triangle diagrams. 
This is a consequence of the fact that we have regulated the trace (\ref{eq:trreg}) in a way that  explicitly 
preserves gauge invariance with respect to the background fields. This corresponds to an asymmetric treatment of
the three legs of the triangle diagram. To restore the symmetry one would have to add local
counterterms, obtaining in this way: 
\be
\label{eq:finright}
A(x) = \frac12 \left[\delta(y) + \delta(y-\pi R)\right] \frac {-1}{96\pi^2} F_{\mu\nu}\tilde F^{\mu\nu} \;. 
\end{equation} 
We have recovered with Fujikawa's approach the gauge anomaly on $S^1/\Z_2$, originally calculated in 
\cite{Arkani-Hamed:2001is} without symmetrization (i.e. they obtained eq.~(\ref{eq:fin})). 
The integrated anomaly is that of a Weyl fermion: the orbifold projection leaves in fact
one unpaired massless state.

The whole discussion can be applied also to the $S^1/(\Z_2 \times \Z^\prime_2)$ case. Now we have to project
as in eq.~(\ref{eq:jacob2}) with respect to both the $\Z_2$'s and the relevant trace turns out to be
\be
\label{eq:Z2'Tr}
A(x) = - \frac12 {\rm Tr}_{\{\psi_n\}}\,(\gamma^5 R + \gamma^5 R') \,.
\end{equation}
The calculation follows precisely the previous one, giving the result (\ref{anomaly})
as obtained in \cite{Scrucca:2001eb}, apart from the already discussed factor $1/3$. 

All the conclusions drawn in the previous sections can be reobtained in this approach. In particular, note that 
Fujikawa's calculation does not tell us anything about the possibility of canceling the anomalous variation
of the fermion measure by local counterterms. As we discussed, in the $S^1/(\Z_2 \times \Z_2^\prime)$ case the 
addition of a heavy fermion with a stepwise mass gives a contribution to the calculation of the 
anomaly that cancels the contribution of the light states. This can be checked in this context: the anomaly
here appears in the measure of integration, so that it does not change if we add a mass for the 
fermion\footnote{In particular it tells us that the anomaly does not depend on the mass profile $m(y)$, 
as it was explicitly checked in ref.~\cite{Arkani-Hamed:2001is} for the $S^1/\Z_2$ case.}. Usually
this does not help to cancel anomalies because no fermion that is allowed by a given symmetry to have a mass
can contribute to the anomaly for that symmetry; here this is not the case.

%%%%%%%%%%
%%%%%%%%%%      References
%%%%%%%%%%

\end{document}